\documentclass[11pt]{article} 
\usepackage{hyperref}

\makeatletter\renewcommand{\section}{\@startsection
{section}{1}{\z@}{-3.5ex plus -1ex minus
    -.2ex}{2.3ex plus .2ex}{\bf }}

\makeatletter\renewcommand{\subsection}{\@startsection{subsection}{2}{\z@}{-3.25ex
plus -1ex minus
   -.2ex}{1.5ex plus .2ex}{\it }}
\makeatletter\renewcommand{\subsubsection}{\@startsection{subsubsection}{3}{-2.45ex}{-3.25ex
plus -1ex minus -.2ex}{1.5ex plus .2ex}{\it }}
\renewcommand{\thesection}{\arabic{section}}

\renewcommand{\theequation}{\thesection.\arabic{equation}}
\makeatletter \@addtoreset{equation}{section}

\usepackage{graphicx}
\usepackage{epsfig} 

\newcommand{\be}{\begin{equation}}
\newcommand{\ee}{\end{equation}}
\newcommand{\bea}{\begin{array}}
\newcommand{\ea}{\end{array}}
\newcommand{\beqa}{\begin{eqnarray}}
\newcommand{\eeqa}{\end{eqnarray}}
\newcommand{\nn}{\nonumber}

\renewenvironment{thebibliography}[1]
     {\baselineskip=16pt plus 2pt minus 1pt
      \section*{\large\refname
        \@mkboth{\MakeUppercase\refname}{\MakeUppercase\refname}}%
     \list{\@biblabel{\@arabic\c@enumiv}}%
           {\settowidth\labelwidth{\@biblabel{#1}}%
            \leftmargin\labelwidth
            \advance\leftmargin\labelsep
            \@openbib@code
            \usecounter{enumiv}%
            \let\p@enumiv\@empty
            \renewcommand\theenumiv{\@arabic\c@enumiv}}%
      \sloppy
      \clubpenalty4000
      \@clubpenalty \clubpenalty
      \widowpenalty4000%
      \sfcode`\.\@m}

\let\fn\footnote
\renewcommand{\footnote}[1]{\linespread{1.1}\fn{#1}\linespread{1.29}}

\usepackage{amsfonts}
\usepackage{amssymb}
\usepackage{mathrsfs}
\usepackage{amsmath,amssymb}
\usepackage{bbm}
\usepackage{bm}
\usepackage{easybmat}

\newcommand{\appendices}{\section*{Appendix}\setcounter{section}{0} \setcounter{equation}{0}
\renewcommand{\thesection}{\Alph{section}.}
\renewcommand{\theequation}{\thesection\arabic{equation}}}

\def\tyng(#1){\hbox{\tiny$\yng(#1)$}}

\begin{document}

\begin{titlepage}
\begin{flushright}
{\bf  \today} 
\end{flushright}

\vskip 2 em

\begin{center}
\centerline{{\Large \bf New Fuzzy Extra Dimensions from $SU({\cal N})$ Gauge Theories}} 

\vskip 3em

\centerline{\large \bf S. K\"{u}rk\c{c}\"{u}o\v{g}lu$^\dagger$}

\vskip 1em

\centerline{\sl $^\dagger$ Middle East Technical University, Department of Physics,}
\centerline{\sl Dumlupinar Boulevard, 06800, \c{C}ankaya, Ankara, Turkey}

\vskip 1em

{\sl kseckin@metu.edu.tr \,} 

\end{center}

\vskip 2 em

\begin{quote}
\begin{center}
{\bf Abstract}
\end{center}
\vskip 1em
We start with an $SU(\cal {N})$ Yang-Mills theory on a manifold ${\cal M}$, suitably coupled to scalar fields in the adjoint representation of $SU({\cal N})$, which are forming a doublet and a triplet, respectively under a global 
$SU(2)$ symmetry. We show that a direct sum of fuzzy spheres $S_F^{2 \, Int} := S_F^2 (\ell) \oplus S_F^2 (\ell) \oplus  S_F^2 \left ( \ell + \frac{1}{2} \right ) \oplus S_F^2 \left ( \ell - \frac{1}{2} \right )$ emerges as the vacuum solution after the spontaneous breaking of the gauge symmetry and lay the way open for us to interpret the spontaneously broken model as a $U(n)$ gauge theory over ${\cal M} \times S_F^{2 \, Int}$. Focusing on a $U(2)$ gauge theory we present complete parameterizations of the $SU(2)$-equivariant, scalar, spinor and vector fields characterizing the effective low energy features of this model. Next, we direct our attention to the monopole bundles $S_F^{2 \, \pm} := S_F^2 (\ell) \oplus S_F^2 \left ( \ell \pm \frac{1}{2} \right )$ over $S_F^2 (\ell)$ with winding numbers $\pm 1$, which naturally come forth through certain projections of $S_F^{2 \, Int}$, and give the parameterizations of the $SU(2)$-equivariant fields of the $U(2)$ gauge theory over ${\cal M} \times S_F^{2 \, \pm}$ as a projected subset of those of the parent model. Making contact to our earlier work \cite{Seckin-Derek}, we explain the essential features of the low energy effective action that ensues from this model after dimensional reduction. Replacing the doublet with a $k$-component multiplet of the global $SU(2)$, we provide a detailed study of vacuum solutions that appear as direct sums of fuzzy spheres as a consequence of the spontaneous breaking of $SU(\cal {N})$ gauge symmetry in these models and obtain a class of winding number $\pm (k-1) \in {\mathbb Z}$ monopole bundles $S_F^{2 \,, \pm (k-1)}$ over $S_F^2 (\ell)$ as certain projections of these vacuum solutions and briefly discuss their equivariant field content. We make the observation that $S_F^{2 \, Int}$ is indeed the bosonic part of the $N=2$ fuzzy supersphere with $OSP(2,2)$ supersymmetry and construct the generators of the $osp(2,2)$ Lie superalgebra in two of its irreducible representations using the matrix content of the vacuum solution $S_F^{2 \, Int}$. 
Finally, we show that our vacuum solutions are stable by demonstrating that, they form mixed states with non-zero von Neumann entropy.
\end{quote}

\end{titlepage}

\setcounter{footnote}{0}

\newpage

\section{Introduction}

Dynamical generation of fuzzy extra dimensions in the form of a fuzzy sphere $S_F^2$ or the product $S_F^2 \times S_F^2$ from $SU({\cal N})$ gauge theories coupled to scalar fields in the adjoint representation of the gauge group \cite{Aschieri:2006uw,Aschieri:2003vy,Steinacker3} (see \cite{Zoupanos-Review} for a review) constitutes recent intriguing examples of the ideas introduced in \cite{ArkaniHamed:2001ca, ArkaniHamed:2001nc}, and known by the name of deconstruction in the literature. In the latter, it was shown that extra dimensions may emerge dynamically in a four-dimensional renormalizable and asymptotically free gauge theory, while in the aforementioned recent studies \cite{Aschieri:2006uw,Steinacker3}, it was demonstrated that vacuum expectation values of the scalar fields form fuzzy sphere(s) and fluctuations around these vacuum configurations take the form of gauge fields over $S_F^2$ or $S_F^2 \times S_F^2$ leading to the interpretation that the emerging theories after spontaneous symmetry breaking are gauge theories over $M^4 \times S_F^2$ or $M^4 \times S_F^2 \times S_F^2$ with smaller gauge symmetry groups. This latter fact is also ascertained by the construction of a tower of Kaluza-Klein (KK) modes of the gauge fields. Inclusion of fermions in models over $M^4 \times S_F^2$ or $M^4 \times S_F^2 \times S_F^2$ have also been investigated in the recent past and it has been found out that, low energy physics obtained from KK modes analysis have "mirror fermions", where chiral fermions come with pairs of opposite chirality and quantum numbers \cite{Steinacker3, Steinacker2}.

These emerging models with fuzzy extra dimensions have connections with effective models arising at the low energy limit of string theories, such as the BMN matrix model \cite{Berenstein:2002jq, Dasgupta:2002hx}, massive deformations of the $N=4$ super Yang-Mills theories; for instance the $N=1^*$ models \cite{Dorey:1999sj,Dorey:2000fc,Auzzi:2008ep} to name a few. In fact, the model investigated in \cite{Steinacker3} has the same field content as the $N = 4$ super Yang-Mills theory, but it is a massive deformation of the latter involving potential terms breaking the SUSY completely and the global $SU(4)$ $R$-symmetry down to a global $SU(2) \times SU(2)$.  Another related paper \cite{Grosse:2010zq} launched an investigation, starting from a higher dimensional $SU({\cal N})$ Yang-Mills matrix model, which is similar to the IKKT matrix model \cite{Ishibashi:1996xs} associated to the low energy physics of the type IIB superstring theory  and considered the spontaneous symmetry breaking schemes mediated by the appearance of fuzzy spheres. They have shown that surviving gauge group after symmetry breaking, which is of the form $SU(3)_c \times SU(2)_L \times U(1)_Q$, couples to all fields of the standard model in a suitable manner and the resulting low energy physics appears to be an extension of the standard model. In \cite{Zoupanos-1} certain orbifold projections of $N=4$ super Yang-Mills theory have been considered and it was shown that utilizing soft supersymmetry breaking terms reveal extra dimensions which are twisted fuzzy spheres consistent with orbifolding. Implications of this model related to Standard model and MSSM at low energies are also studied in \cite{Zoupanos-1}. Other related new results have also been reported in \cite{Steinacker:2014lma, Steinacker:2014eua}.  

In our recent work, we have given the equivariant parameterizations of $U(2)$ and $U(4)$ gauge theories over ${\cal M} \times S_F^2$ and  ${\cal M} \times S_F^2 \times S_F^2$, respectively, which has provided further insights on the structure of these theories that characterize their low energy physics \cite{Seckin-Derek, Seckin1, Seckin2}. In these studies, we have adapted and employed coset space dimensional reduction (CSDR) techniques discussed in \cite{Forgacs, Zoupanos, Zoupanos-Review} (See also, \cite{Aschieri:2003vy} in this context). Essential idea beneath this technique may be presented briefly by considering a Yang-Mills theory with a gauge group $S$ over the product space ${\cal M} \times G/H$. Group $G$ has a natural action on its coset, and demanding that the Yang-Mills gauge fields be invariant under this $G$ action up to $S$ gauge transformations leads immediately to $G$-equivariant parametrization of the gauge fields. Subsequently, such models may be dimensionally reduced by integrating over the coset space $G/H$ and explicit form of the low energy effective action may be obtained. After, obtaining the $SU(2)$ and $SU(2) \times SU(2)$ equivariant parameterizations of fields in \cite{Seckin-Derek, Seckin2}, we were able to compute the dimensionally reduced actions by tracing over the fuzzy spheres, and found that Abelian Higgs type models with one or several (four for the case in \cite{Seckin2}) complex scalar fields and additionally some real scalars, which has attractive or repulsive (multi)vortex solutions depending on the couplings of the scalars and the gauge fields in the parent $SU({\cal N})$ theory, emerge. The case of ${\cal M} = {\mathbb R}_\theta$, the Moyal plane, was treated in \cite{Seckin1} and, we have found noncommutative vortices and flux tube solutions in the low energy limit. Other recent related work on equivariant reduction over extra dimensions include \cite{Popov-Szabo, Lechtenfeld1, Popov, Popov2, Dolan-Szabo, Landi-Szabo,Lechtenfeld:2003cq}.

It is also worthwhile to remark that, results that bear resemblance especially to our findings in \cite{Seckin-Derek, Seckin2}, have also emerged in the context of ABJM models \cite{Aharony:2008ug, Nastase:2009ny}. The latter are, as well-known, $N=6$ SUSY $U({\cal N}) \times U({\cal N})$ Chern-Simons gauge theories at the level $(k, -k)$ with scalar and spinor fields in the bifundamental and fundamental representation, respectively of the $SU(4)$ $R$-symmetry group. A particular massive deformation of the ABJM model \cite{Gomis:2008vc,Terashima:2008sy} preserving all the supersymmetry but partially breaking the $R$ symmetry down to $SU(2) \times SU(2) \times U(1)_A \times U(1)_B \times {\mathbb Z}_2$ leads to vacuum solutions of the model, which are fuzzy sphere(s) in the bifundamental formulation realized in terms of the GRVV matrices \cite{Gomis:2008vc, Nastase:2010uy}. A particular parametrization of the fields given in \cite{Mohammed:2012gi,Mohammed:2012rd} leads to a low energy effective action involving four complex scalar fields interacting with a sextic potential, containing the relativistic Landau-Ginzburg model in a certain limit.

These developments indicate that there is ample motivation for further exploring the structure of gauge theories with spontaneously generated fuzzy extra dimensions. In this article we find a new class of fuzzy extra dimensions emerging from an $SU(\cal {N})$ gauge theory as direct sums of fuzzy spheres. Specifically, we orient the developments starting with an $SU(\cal {N})$ Yang-Mills theory on a manifold ${\cal M}$, suitably coupled to two separate sets of scalar fields both in the adjoint representation of $SU({\cal N})$, which are forming a doublet and a triplet under the global $SU(2)$ symmetry. Although, we only admit the bilinears (or composites) of the $SU(2)$-doublets that transform as a vector under the global $SU(2)$, we are able to detect various new features in the model, which can be ascribed to the presence of the doublet fields. We find that a direct sum of fuzzy spheres $S_F^{2 \, Int} := S_F^2 (\ell) \oplus S_F^2 (\ell) \oplus  S_F^2 \left ( \ell + \frac{1}{2} \right ) \oplus S_F^2 \left ( \ell - \frac{1}{2} \right )$ appears as fuzzy extra dimensions after the spontaneous breaking of the gauge symmetry and forms the vacuum configuration of our model. By considering the fluctuations around this vacuum, we show that the spontaneously broken model may be interpreted as a $U(n)$ gauge theory over ${\cal M} \times S_F^{2 \, Int}$. In order to place this interpretation on a firmer ground we focus on the $U(2)$ theory and present complete parameterizations of the $SU(2)$-equivariant, scalar, spinor and vector fields characterizing the effective low energy structure of this model. Strikingly, we encounter the equivariant spinor fields as a direct consequence of (although, implicitly in the form of bilinears) admitting $SU(2)$-doublets.

We note that monopole bundles $S_F^{2 \, \pm} := S_F^2 (\ell) \oplus S_F^2 \left ( \ell \pm \frac{1}{2} \right )$ over $S_F^2 (\ell)$ \cite{Book, Steinacker:2003sd, Aoki:2003ye}with winding numbers $\pm 1$, naturally appear after a certain projection of $S_F^{2 \, Int}$, which we identify and subsequently give the parameterizations of the $SU(2)$-equivariant fields of the $U(2)$ theory over ${\cal M} \times S_F^{2 \, \pm}$ as a projected subset of those on ${\cal M} \times S_F^{2 \, Int}$. We make the observation that the low energy effective action that ensues from this model by tracing over (dimensionally reducing) $S_F^{2 \, Int}$ may be seen as two decoupled abelian Higgs type models by making a comparison with the results of our earlier work \cite{Seckin-Derek}. 

Replacing the two-component spinors with a $k$-component multiplet of the global $SU(2)$ and admitting them in our model only through their bilinears we find vacuum solutions, which are given as particular direct sums of fuzzy spheres. In section 4, we inspect these models in considerable detail and determine the aforementioned vacuum solutions and discuss their equivariant field content for the cases of $k=3$ and $k=4$. In addition, we obtain a particular class of winding number $\pm (k-1) \in {\mathbb Z}$ monopole bundles $S_F^{2 \,, \pm (k-1)}$ as certain projections of these vacuum solutions. 

An intriguing result that we came across in  our studies is that the vacuum configuration $S_F^{2 \, Int}$ forms the bosonic part of the $N=2$ fuzzy supersphere with $OSP(2,2)$ supersymmetry \cite{Book, Balachandran:2002jf, Kurkcuoglu:2003ke, Hasebe}. This follows from a comparison of the direct sum of $SU(2)$ IRRs that is used to describe $S_F^{2 \, Int}$ and the $SU(2)$ IRR decomposition of the typical superspin IRRs of $OSP(2,2)$. Moreover, we manage to use the matrix content of the vacuum solution $S_F^{2 \, Int}$ to give a construction of the generators of $OSP(2,2)$ in its $3$-dimensional atypical and the $4$-dimensional typical irreducible representation. 

We discuss the stability of our vacuum solutions using the recent novel approach developed in \cite{Acharyya:2014nfa} which addresses the mixed state nature of configurations with several fuzzy spheres and their quantum entropy, relying on the broader considerations of quantum entropy and its ambiguities recently discussed in  \cite{Balachandran:2012pa, Balachandran:2013kia}. We show that, our vacuum configurations which are direct sums of fuzzy spheres, do indeed form mixed states with non-zero von Neumann entropy, while single fuzzy sphere solutions form pure states with vanishing entropy. Stability of our vacuum solutions follows, since mixed states can not go to pure states under unitary evolution. A detailed account on this is provided in section $6$.

\section{Gauge Theory over ${\cal M} \times S_F^{2 \, Int}$}

\subsection{The Model}

We consider the following $SU(\cal{N})$ Yang-Mills theory with the action
\be
S = \int_{{\cal M} } \, \mbox{Tr}_{{\cal N}} \Big (
\frac{1}{4g^2} F_{\mu \nu}^\dagger F^{\mu \nu} + 
(D_\mu \Phi_a)^\dagger (D^\mu \Phi_a) \Big ) +
\frac{1}{\tilde{g}^2}  V(\Phi_a) \,,
\label{eq:actionfirstinternal}
\ee
where
\be
V(\Phi_a) =  \mbox{Tr}_{{\cal N}} \big ( F_{ab}^\dagger F_{ab} \big ) \,.
\label{eq:pot}
\ee
In (\ref{eq:actionfirstinternal}), $F_{\mu \nu}$ is the curvature associated to the $su({\cal N})$ valued connection $A_\mu$. 
We take $A_\mu$ anti-Hermitian ($A_\mu^\dagger = - A_\mu$) and $\Phi_a \,(a=1,2,3) \in Mat({\cal N})$ are anti-Hermitian ($\Phi_a^\dagger = - \Phi_a$) scalar fields, transforming in the adjoint representation of ${\rm SU(}{\cal N}{\rm )}$ as 
\be
\Phi_a \rightarrow U^\dagger \Phi_a U \,, \quad U \in {\rm SU(}{\cal N}{\rm )} \,,
\ee
and in the vector representation of the global $SO(3)\simeq SU(2)$ symmetry of the action. The covariant derivative of $\Phi_a$ is
\be
D_\mu \Phi_a  = \partial_\mu \Phi_a + \lbrack A_\mu \,, \Phi_a \rbrack \,.
\ee 
With a hindsight of future developments, the quantity $F_{ab}$ is defined as 
\be
F_{ab} := \lbrack \Phi_a \,, \Phi_b \rbrack - \varepsilon_{abc} \Phi_c \,.
\label{eq:curvaturefuzzy}
\ee
In (\ref{eq:actionfirstinternal}) and (\ref{eq:pot}) $g$ and $\tilde{g}$ are the coupling constants and $\mbox{Tr}_{{\cal N}} = {\cal N}^{-1} \mbox{Tr}$ denotes the normalized trace. 

We assume that the matrices $\Phi_a \,(a=1,2,3)$ have the following structure:
\be
\Phi_a = \phi_a + \Gamma_a \,, \quad \Gamma_a = -\frac{i}{2}\Psi^\dagger {\tilde \tau}_a \Psi  \,,
\label{eq:isospinor1}
\ee
where
\be
\Psi = \left (
\begin{array}{c}
\Psi_1 \\
\Psi_2
\end{array}
\right ) \,,
\label{eq:isospinor2}
\ee
is a doublet of the global $SU(2)$ and $\phi_a$, $\Psi_\alpha \in Mat({\cal N})$ ($\alpha = 1, 2$) are anti-Hermitian and transform adjointly under the $SU(\cal{N})$ as $\phi_a \rightarrow U^\dagger \phi_a U$ and $\Psi_\alpha \rightarrow U^\dagger \Psi_\alpha U$. Clearly $\Gamma_a$'s are also anti-Hermitian and transform adjointly 
\be
\Gamma_a \rightarrow U^\dagger \Gamma_a U \,, \quad U \in {\rm SU(}{\cal N}{\rm )} \,,
\ee
under ${\rm SU(}{\cal N}{\rm )}$ and transforms in the vector of the global $SU(2)$. In (\ref{eq:isospinor1}), ${\tilde \tau}_a$ stands for $\tau_a \otimes 1_{{\cal N}}$, $\tau_a$ being the Pauli matrices. In our model we only admit the bilinears $\Gamma_a$'s of the fields $\Psi$, but as we shall see many new features emerge, which can be ascribed to introducing the latter in our model. 

This theory spontaneously develops extra dimensions in the form of direct sums of fuzzy spheres with many novel features as we demonstrate next. 

We will consider the generalisation of (\ref{eq:isospinor2}) to $k$-component multiplets transforming under the $k$-dimensional IRR of $SU(2)$ and their implications in section $4$.

\subsection{The Vacuum Structure and Gauge Theory over ${\cal M} \times S_F^{2 \, Int}$ }

We observe that $V(\Phi_a)$ is positive definite, and it is minimized by the solutions  of
\be
F_{ab} = 0 \,, \quad 
\label{eq:mini1}
\ee
Solutions of this equation have been discussed previously \cite{Berenstein:2002jq, Dorey:1999sj, Dorey:2000fc, Aschieri:2006uw}. In general, they are given in terms of ${\cal N} \times {\cal N}$ matrices carrying direct sums of irreducible representations of $SU(2)$. In the present case, we require that $\Gamma_a$'s are bilinear in $\Psi$ as introduced in (\ref{eq:isospinor2}) and it is not possible to pick $\Gamma_a$ in an arbitrary IRR of $SU(2)$, as the corresponding $\Psi$ will not exist in general. We restrict ourselves to a possible solution for which neither $\phi_a$ nor $\Gamma_a$ vanishes. Assuming that the dimension ${\cal N}$ of the matrices $\Phi_a$ factorizes as $4 (2 \ell +1) n$, (\ref{eq:mini1}) is solved by configurations of the form 
\be
\Phi_a = ( X_a^{(2 \ell + 1)} \otimes {\bf 1}_4 \otimes {\bf 1}_n ) + ( {\bf 1}_{2 \ell + 1} \otimes \Gamma_a^0 \otimes {\bf 1}_n )  \,,
\label{eq:minisol}
\ee
with
\be
\lbrack X_a \,, X_b \rbrack = \varepsilon_{abc} X_c \,, \quad \lbrack \Gamma_a^0 \,, \Gamma_b^0 \rbrack = \varepsilon_{abc} \Gamma_c^0 \,,
\ee
up to gauge transformations $\Phi_a \rightarrow U^\dagger \Phi_a U$. $X_a^{(2 \ell + 1)}$ are the (anti-Hermitian) generators of ${\rm SU(2)}$ in the irreducible representation $\ell$ and 
\be
\Gamma_a^0 = - \frac{i}{2} \psi^\dagger \tau_a \psi \,,
\ee
are $4 \times 4$ matrices carrying a reducible representation of ${\rm SU(2)}$.  To facilitate the developments, it is necessary to describe the structure of the latter.

We introduce two sets of fermionic annihilation-creation operators, fulfilling the anti-commutation relations
\be
\lbrace b_\alpha \,, b_\beta \rbrace = 0 \,, \quad \lbrace b_\alpha^\dagger \,, b_\beta^\dagger \rbrace = 0 \,, \quad \lbrace b_\alpha \,, b_\beta^\dagger  \rbrace = \delta_{\alpha \beta} \,.
\ee
They span the $4$-dimensional Hilbert space with the basis vectors
\be
|n_1 \,, n_2 \rangle \equiv (b_1^\dagger)^{n_1} (b_2^\dagger)^{n_2} | 0 \,, 0 \rangle \,, \quad n_1 \,, n_2 = 0 \,,1 \,. 
\label{eq:basis1}
\ee
Taking the two-component spinor 
\be
\psi = \left (
\begin{array}{c}
\psi_1 \\
\psi_2
\end{array}
\right )
:=
\left (
\begin{array}{c}
b_1 \\
b_2
\end{array}
\right ) \,,
\label{solspi}
\ee
it is easy to see that $\Gamma_a^0$'s fulfill the ${\rm SU(2)}$ commutation relations and $b_\alpha$, $b_\alpha^\dagger$ are ${\rm SU(2)}$ spinors:
\be
\lbrack b_\alpha \,, \Gamma_a^0 \rbrack = - \frac{i}{2} (\tau_a)_{\alpha \beta} b_\beta \,, \quad \lbrack b_\alpha^\dagger \,, \Gamma_a^0 \rbrack =  \frac{i}{2} (\tau_a)_{\beta \alpha} b_\beta^\dagger \,.
\ee
$\Gamma_a^0$ furnish a reducible representation of ${\rm SU(2)}$ composed of two inequivalent singlets and a doublet, i.e. it has the irreducible decomposition 
\be
0_{\bm 0} \oplus 0_{\bm 2} \oplus \frac{1}{2} \,.
\label{eq:irrdec}
\ee
Here the inequivalent singlets are distinguished by the eigenvalue of $N = N_1 + N_2$. With the notation of (\ref{eq:basis1}) the singlets states are $|0\,,0 \rangle$ and $|1 \,,1 \rangle$ and carry the eigenvalues of $N$, which are $0$ and $2$ respectively and they are denoted by the subscripts appearing in (\ref{eq:irrdec}). 

The quadratic Casimir operator $(\Gamma_a^0)^2$ can be expressed as 
\be
(\Gamma_a^0)^2 = - \frac{3}{4} N + \frac{3}{2} N_1 N_2 \,, \quad N_1 = b_1^\dagger b_1 \,, \quad N_2 = b_2^\dagger b_2 \,, \quad N= N_1 + N_2 \,,
\ee
which of course has the eigenvalue $0$ on the singlets and $-\frac{3}{4}$ on the doublet. It also follows from the anti-commutators of the oscillators that $N_1$ and $N_2$ are projectors:
\be
N_1^2 = N_1 \,, \quad N_2^2 = N_2 \,.
\ee
We can define the projections to the singlet and doublet subspaces respectively as
\be
\begin{aligned}
P_0 &= \frac{(\Gamma_a^0)^2 + \frac{3}{4}}{\frac{3}{4}}  = 1 - N + 2 N_1 N_2 \,, \\
P_{\frac{1}{2}} &= - \frac{(\Gamma_a^0)^2 }{\frac{3}{4}} = N - 2 N_1 N_2  \,.
\end{aligned}
\ee
We can split $P_0$ into two projectors corresponding to two inequivalent singlet representations $0_{\bm 0}$ and $0_{\bm 2}$ as
\be
\begin{aligned}
P_{0_{\bm 0}} &= - \frac{1}{2} (N - 2 ) P_0 = 1 - N + N_1 N_2 \,,  \\
P_{0_{\bm 2}} &= \frac{1}{2} N P_0 =  N_1 N_2 =  - \frac{1}{2} N +\frac{1}{2} P_{\frac{1}{2}} \,.
\end{aligned}
\ee
$\Gamma_a^0$ also fulfill 
\be
\Gamma_a^0 \Gamma_b^0 = - \frac{1}{4} \delta_{ab} P_{\frac{1}{2}} + \frac{1}{2}\varepsilon_{abc} \Gamma_c^0 \,, \quad \mbox{Tr} \Gamma_a^0 \Gamma_a^0 = - \frac{3}{2}.
\ee
We relegate some useful identities involving $\Gamma_a^0$ and further related formulas to an appendix and continue our discussion. 

Going back now to the vacuum configuration (\ref{eq:minisol}), we observe that its $SU(2)$ IRR content follows from the Clebsch-Gordan decomposition as
\be
\ell \otimes \left (0_{\bm 0} \oplus 0_{\bm 2} \oplus \frac{1}{2} \right) \equiv  \ell \oplus \ell \oplus \left ( \ell + \frac{1}{2} \right )\oplus \left ( \ell - \frac{1}{2} \right )\,, \quad \ell \neq 0 \,.
\label{eq:1deco1}
\ee
Let us introduce the short-hand notation 
\be
(X_a^{(2 \ell + 1)} \otimes {\bf 1}_4 \otimes {\bf 1}_n) + ( {\bf 1}_{2 \ell + 1} \otimes \Gamma_a^0 \otimes {\bf 1}_n ) =: X_a +\Gamma_a^0 =: D_a
\label{eq:config2}
\ee 
A unitary transformation $U^\dagger D_a U$ can bring $D_a$ to the block diagonal form
\be
{\cal D}_a := U^\dagger D_a U =  (X_a^{(2\ell+1)} \,, X_a^{(2 \ell+1)} \,,X_a^{(2 \ell+2)} \,, X_a^{(2 \ell)} ) \otimes {\bm 1}_{n} \,,
\label{eq:calDa}
\ee
with 
\begin{multline}
{\cal D}_a {\cal D}_a = \mbox{Diag} \Big (-\ell(\ell +1) {\bm 1}_{(2\ell+1)n} \,, -\ell(\ell +1) {\bm 1}_{(2\ell+1)n} \,, - (\ell +\frac{1}{2})(\ell + \frac{3}{2}) {\bm 1}_{(2\ell+2)n} \,, \\
- (\ell - \frac{1}{2})(\ell + \frac{1}{2}) {\bm 1}_{(2\ell)n} \Big) \,.
\end{multline}
Thus, we see that we can interpret the vacuum configuration for $\Phi_a$ as a direct sum of four concentric fuzzy spheres
\be
S_F^{2 \, Int} := S_F^2 (\ell) \oplus S_F^2 (\ell) \oplus  S_F^2 \left ( \ell + \frac{1}{2} \right ) \oplus S_F^2 \left ( \ell - \frac{1}{2} \right ) \,.
\label{eq:sfint}
\ee
Level of all four fuzzy spheres are correlated by the the parameter $\ell$. This internal structure of the vacuum is well reflected by the derivations on $S_F^{2 \, Int}$ that we introduce in (\ref{eq:fab}). In fact as we shall see in section 5, this vacuum structure fits just right to the superspin $j$ IRR of the supergroup $OSP(2,2)$. For this reason, we may think of the vacuum as the even part of a $N = 2$ fuzzy supersphere \cite{Balachandran:2002jf, Book, Hasebe}.

Now, the configuration in (\ref{eq:minisol}) spontaneously breaks the ${\rm SU}({\cal N})$ down to ${\rm U}(n)$ which is the commutant of $\Phi_a$ in (\ref{eq:minisol}). The global $SU(2)$ is spontaneously broken completely by the vacuum. There is however, a combined global rotation and a gauge transformation under which the vacuum remains invariant. 

Fluctuations about this vacuum may be written as
\be
\Phi_a = X_a + \Gamma_a^0 + A_a =: D_a + A_a \,,
\label{eq:config1} 
\ee
where $A_a \in u(4) \otimes u(2\ell+1)\otimes u(n)$. 

We may interpret $A_a$ $(a=1,2,3)$ as the three components of a ${\rm U}(n)$ gauge field on $S_F^{2 \, Int}$. $\Phi_a$ are indeed the ``covariant coordinates'' on $S_F^{2 \, Int}$ and $F_{ab}$ is the field strength, which takes the form
\be
\begin{aligned}
F_{ab} &= \lbrack X_a + \Gamma_a^0 \,, A_b \rbrack - \lbrack X_b +\Gamma_b^0 \,, A_a \rbrack + \lbrack A_a \,, A_b \rbrack - \varepsilon_{abc} A_c \,, \\
&= \lbrack D_a \,, A_b \rbrack - \lbrack D_b \,, A_a \rbrack + \lbrack A_a \,, A_b \rbrack - \varepsilon_{abc} A_c  \,,
\end{aligned}
\label{eq:fab}
\ee
when expressed in terms of the gauge fields $A_a$. We also note that in the second line above we have used $\mbox{ad} D_a \cdot = \lbrack D_a \,, \cdot \rbrack$, which are the natural derivations on $S_F^{2 \, Int}$. 

To summarize, with (\ref{eq:config1}) the action in (\ref{eq:actionfirstinternal}) takes the form of a ${\rm U(n)}$ gauge theory\footnote
{In fact, the gauge fields are in general valued in the enveloping algebra ${\cal U}(n)$ of $u(n)$. This is a well-known feature of non-commutative field theories \cite{Madore:2000en, Jurco:2000ja}. This fact, will be more apparently seen, when we give the equivariant parameterizations of the gauge fields in section 3. The latter involve intertwiners of the IRRs of $su(2)$, which are elements of the enveloping algebra ${\cal SU}(2)$.} on ${\cal M} \times S_F^{2 \, Int}$ with the gauge field components $A_M = (A_\mu \,, A_a) \in u(2\ell+1) \otimes u(4) \otimes u(n)$ and field strength tensor 
$F_{MN} = (F_{\mu \nu} \,, F_{\mu a} \,, F_{ab})$ where
\begin{eqnarray}
F_{\mu\nu} &=& \partial_\mu A_\nu - \partial_\nu A_\mu + [A_\mu,A_\nu]  \nn \\
F_{\mu a} &=& D_\mu \Phi_a = \partial_\mu A_a - \lbrack X_a + \Gamma_a^0 \,, A_\mu \rbrack + \lbrack A_\mu \,, A_a \rbrack \,,  \\
F_{ab} &=& \lbrack X_a + \Gamma_a^0 \,, A_b \rbrack - \lbrack X_b +\Gamma_b^0 \,, A_a \rbrack + \lbrack A_a \,, A_b \rbrack - \varepsilon_{abc} A_c  \,. \nn
\end{eqnarray}

It is important to remark here that for gauge theories on fuzzy spaces, there is no canonical way to separate the component of the fuzzy gauge field normal to the fuzzy sphere(s). This is usually achieved by imposing a gauge invariant condition, which disentangles the normal component in the commutative limit $\ell \rightarrow \infty$ \cite{Karabali:2001te,Balachandran:2003ay, Book}, or by turning the normal component into a scalar field with a large mass and adding it to the action by a Lagrange multiplier like term \cite{Steinacker:2003sd, Aoki:2003ye}. Here, we have admitted a vacuum solution of concentric fuzzy spheres carrying the direct sum representation (\ref{eq:1deco1}) and therefore as discussed in \cite{Aschieri:2006uw} the latter choice can not be availed. Following \cite{Karabali:2001te, Balachandran:2003ay, Book} we consider imposing the constraints
\be
(X_a + \Gamma_a + A_a)^2 = (X_a + \Gamma_a)^2 = - (\ell + \gamma)(\ell+\gamma+1) {\bm 1}
\ee
where $\gamma$ is taking on the values $\pm \frac{1}{2} \,, 0$. In the commutative limit $\ell \rightarrow \infty$, we see that this condition gives the transversality condition on $\Gamma_a + A_a$ as ${\hat x_a} (\Gamma_a + A_a) \longrightarrow -\gamma$, as long as $A_a$ are smooth and bounded for $\ell \rightarrow \infty$ and therefore converges to the commutative field $A_a(x)$ in this limit. Here ${\hat x}_a$ with ${\hat x}_a {\hat x}_a =1$ are the coordinates on the sphere $S^2$ and we have used the fact that $\frac{X_a}{\ell} \longrightarrow {\hat x}_a$ when $\ell \rightarrow \infty$.

It is possible to elaborate on the emergence of such a gauge theory with fuzzy extra dimensions, by working out the Kaluza-Klein (KK) tower of states on ${\cal M}$ due to the extra dimensions $S_F^{2 \, Int}$ in a manner similar to that given in \cite{Aschieri:2006uw} for fuzzy extra dimensions in the form of a $S_F^{2}$ and $S_F^{2}$ with nonzero monopole number. The latter lead to a KK spectrum with ground states separated from the rest of excitations by large energy gaps. In the case of $S_F^{2}$ the ground state of the KK tower is gapless and the resulting low energy effective action (LEA) is that of $U(n)$ Yang-Mills on ${\cal M}$. As for the latter, the off-diagonal ground state KK modes acquire masses, while the diagonal ones remain massless with the LEA differing from the former by a constant additive term proportional to the square of the monopole winding number. In the present case, it is reasonable to expect that a similar KK structure to occur, corroborating with the emergence of the ${\rm U(n)}$ gauge theory on ${\cal M} \times S_F^{2 \, Int}$. However, we are not going to direct our developments in this way, but will focus on the formulation of equivariant gauge fields for $U(2)$ theory and draw qualitative conclusions for the low energy physics emerging from such equivariant gauge fields. 

\subsection{Projection to the Monopole Sectors}

Another highly interesting structure that emerges from $S_F^{2 \, Int}$, is the projection of $S_F^{2 \, Int}$ to
\be
S_F^{2 \, \pm} := S_F^2 (\ell) \oplus S_F^2 \left ( \ell \pm \frac{1}{2} \right ) \,,
\label{eq:Mon1}
\ee
which may readily be interpreted as the monopole bundles over $S_F^2 (\ell)$ with winding numbers $\pm 1$ \cite{Steinacker:2003sd, Book}.

Let us start with the projector
\be
\Pi_\alpha = \prod_{\beta \neq \alpha} \frac{-(X_a + \Gamma^0_a)^2 - \lambda_\beta(\lambda_\beta +1){\bm 1}}{\lambda_\alpha(\lambda_\alpha +1) - \lambda_\beta(\lambda_\beta +1)} \,,
\label{eq:proo1}
\ee
where $\alpha = 0 \,, + \,, -$ and $\lambda_\alpha$ take on the values $\ell$, $\ell+\frac{1}{2}$ and $\ell-\frac{1}{2}$ respectively. $\Pi_\alpha$'s project to the irreducible subspaces with the IRR content $\ell \oplus \ell$, $\ell+\frac{1}{2}$ and $\ell-\frac{1}{2}$. We see that the projection $\Pi_0$ may be written as
\be
\Pi_0 = {\bf 1}_{2 \ell + 1} \otimes P_0 \otimes {\bm 1}_n
\ee
as a short calculation can demonstrate and therefore we may further construct
\be
\Pi_{0_{\bm 0}}:= {\bf 1}_{2 \ell + 1} \otimes P_{0_{\bm 0}} \otimes {\bm 1}_n \,, \quad \Pi_{0_{\bm 2}}:= {\bf 1}_{2 \ell + 1} \otimes  P_{0_{\bm 2}} \otimes {\bm 1}_n \,, \quad \Pi_0 = \Pi_{0_{\bm 0}} + \Pi_{0_{\bm 2}} \,, 
\ee
as projections to the subspaces with the occupation numbers $N=0$ and $N=2$, respectively. We also note that we may write
\be
\Pi_\frac{1}{2} : = \Pi_+ + \Pi_- = {\bf 1}_{2 \ell + 1} \otimes P_{\frac{1}{2}} \otimes {\bm 1}_n \,.
\ee

Projection from $S_F^{Int}$ in (\ref{eq:sfint}) onto the monopole bundle $S_F^\pm$ in (\ref{eq:Mon1}) is facilitated by either of the projectors 
\be 
(1 - \Pi_\mp)(1 - \Pi_{0_{\bm 0}}) \,, \quad  (1 - \Pi_\mp)(1 - \Pi_{0_{\bm 2}}) \,.
\ee

Monopole sectors with winding numbers $\pm 1$ over fuzzy sphere were found as possible vacuum solutions in the model treated in \cite{Aschieri:2006uw} in which only an adjoint triplet of scalar fields $\phi_a$ were present. In our model however, appearance of the monopole sectors can be attributed to the presence the doublet $\Psi$ transforming under the fundamental IRR of the global $SU(2)$. This allows us to write down the equivariant parametrization of gauge fields in a suitable manner as we shall see in the ensuing sections and it naturally leads to the presence of equivariant spinor fields which do not appear otherwise. In addition to these, generalization of the doublet field to all possible higher dimensional multiplets enables us to give a systematic treatment of a family of fuzzy monopole bundles with winding numbers $m \in{\mathbb Z}$ appearing as fuzzy extra dimensions. This is discussed in section $4$ as we have already noted before.

To keep track of different projections appearing in our discussions and to orient the ensuing developments we list the projections $\Pi_k \in \mbox{Mat} ((2\ell+1)4n)$ ($k:= 0, \frac{1}{2}, 0_{\bm 0}, 0_{\bm 2}, +, -$) introduced in this section together with the subspaces they project to, in the table (\ref{table:t1})
\begin{table}
\begin{center}
    \begin{tabular}{c | c }
    Projector & To the Representation \\ \hline
    $ \Pi_0$ & $\ell \oplus \ell$ \\
    $\Pi_\frac{1}{2} $ & $(\ell +\frac{1}{2}) \oplus (\ell - \frac{1}{2})$ \\  
    $\Pi_{0_{\bm 0}}$ & $\ell$ \\
    $\Pi_{0_{\bm 2}}$ & $\ell$ \\
    $\Pi_+ = \frac{1}{2} (i Q_I + \Pi_{\frac{1}{2}})$ & $(\ell + \frac{1}{2})$ \\
    $\Pi_- = \frac{1}{2} (-i Q_I + \Pi_{\frac{1}{2}})$ & $(\ell - \frac{1}{2})$ 
\end{tabular}
\caption{Projections $\Pi_k$ and the representations to which they project.}
\vspace{-0.4cm}
\label{table:t1}
\end{center}
\end{table}
Here we have introduced
\be
Q_I = i \frac{X_a \otimes \Gamma_a^0 \otimes {\bm 1}_n -  \frac{1}{4} \Pi_{\frac{1}{2}}}{\frac{1}{2}(\ell + \frac{1}{2})} \,, \quad Q_I^2 = - \Pi_{\frac{1}{2}} \,.
\label{eq:calqi}
\ee

\section{Equivariant Parametrization of $U(2)$ Gauge Fields over ${\cal M} \times S_F^{2 \, Int}$}

We now focus on a ${\rm U(2)}$ gauge theory on ${\cal M} \times S_F^{2 \, Int}$. We are going to obtain the ${\rm SU(2)}$-equivariant parametrizations of gauge fields in the most general setting first to shed some further light to the structure of gauge theories over $S_F^{2 \, Int}$ and subsequently restrict our attention to the monopole sector $S_F^{2 \, \pm}$ given in (\ref{eq:Mon1}). 

Construction of ${\rm SU(2)}$-equivariant gauge fields on $S_F^{2 \, Int}$ can be performed following the ideas in \cite{Seckin-Derek}. We pick a set of symmetry generators $\omega_a$ which generate $SU(2)$ rotations of $S_F^{2 \, Int}$ up to $SU(2)$ gauge transformations. Our choice is
\beqa
\omega_a &=& (X_a^{(2 \ell + 1)} \otimes {\bf 1}_4 \otimes {\bf 1}_2) + ({\bf 1}_{2 \ell + 1} \otimes \Gamma_a^0 \otimes {\bf 1}_2)
- ({\bf 1}_{2 \ell + 1} \otimes {\bf 1}_4 \otimes i \frac{\sigma^a}{2} )  \nn \\
\quad &=:& X_a + \Gamma_a^0 - i \frac{\sigma^a}{2} \\
&=& D_a - i \frac{\sigma^a}{2} \,, \quad  \quad \omega_a \in u(2\ell+1) \otimes u(4) \otimes u(2) \nn
\eeqa
with the consistency condition 
\be
\lbrack \omega_a, \omega_b \rbrack = \varepsilon_{abc} \omega_c \,,
\ee 
which is readily satisfied as can easily be checked.  

$\omega_a$ has the ${\rm SU(2)}$ IRR content
\be
\begin{aligned}
\ell \otimes \left (0_{\bm 0} \oplus 0_{\bm 2} \oplus \frac{1}{2} \right ) \otimes \frac{1}{2} & \equiv \left ({\bm 2} \ell \oplus \left (\ell + \frac{1}{2} \right ) \oplus \left (\ell - \frac{1}{2} \right ) \right ) \otimes \frac{1}{2} 
\label{gp} \\
& \equiv  {\bm 2}  \left (\ell + \frac{1}{2} \right ) \oplus  {\bm 2}  \left (\ell - \frac{1}{2} \right ) \oplus (\ell + 1)\oplus  {\bm 2}  \ell \oplus (\ell - 1)\,, 
\end{aligned} 
\ee
where the bold coefficients stood for the multiplicities of the respective IRRs.

${\rm SU(2)}$-equivariance of the gauge theory on ${\cal M} \times S_F^{2 \, Int}$ requires the fulfillment of the following symmetry constraints,
\be
\begin{aligned}
\lbrack \omega_a \,, A_\mu \rbrack  &=  0  \,,  \\
\lbrack \omega_a \,, \psi_\alpha \rbrack &=  \frac{i}{2} ({\tilde \tau}_a)_{\alpha \beta} \psi_\beta \,, \\
\lbrack \omega_a \,, \phi_b \rbrack  &=  \epsilon_{abc} \phi_c \,. 
\label{eq:cons}
\end{aligned}
\ee

We can determine dimensions of the solution spaces for $A_\mu$, $\psi_\alpha$ and $A_a$ by working out the Clebsch-Gordan decomposition of the adjoint action of $\omega_a$. Part of the Clebsh-Gordan series of interest to us reads
\be
\begin{aligned}
\Bigg ({\bm 2}  \left (\ell + \frac{1}{2} \right ) \oplus {\bm 2} \left (\ell - \frac{1}{2} \right ) \oplus (\ell + 1)\oplus  {\bm 2}  \ell \oplus (\ell - 1)\Bigg ) 
\otimes \Bigg ({\bm 2}  \left (\ell + \frac{1}{2} \right ) \oplus {\bm 2} \left (\ell - \frac{1}{2} \right ) \\
\oplus (\ell + 1)\oplus  {\bm 2}  \ell \oplus (\ell - 1) \Bigg )
\equiv \bm{14} \, \, 0 \oplus \bm {24} \, \, \frac{1}{2} \oplus \bm {30} \, \, 1 \oplus \cdots \,.
\end{aligned}
\label{cgse1}
\ee
We note that the appearance of equivariant spinors in this decomposition is purely due to the fact that we have admitted the doublet field $\Psi$ in our model. We will give the construction of these equivariant spinors shortly.

Correspondence of projections $\Pi_k \in \mbox{Mat} ((2\ell+1) \times 4 \times 2)$ ($k:= 0, \frac{1}{2}, 0_{\bm 0}, 0_{\bm 2}, +, -$) to the representations occurring in \eqref{gp} are listed in table (\ref{table:t2}).

\begin{table}
\begin{center}
    \begin{tabular}{c | c }
    Projector & To the Representation \\ \hline
    $\Pi_0$ & ${\bm 2} (\ell +\frac{1}{2}) \oplus {\bm 2} (\ell - \frac{1}{2})$ \\
    $\Pi_{\frac{1}{2}}$ & $(\ell + 1)\oplus  {\bm 2}  \ell \oplus (\ell - 1)$ \\  
    $\Pi_{0_{\bm 0}} $ & $ (\ell +\frac{1}{2}) \oplus  (\ell - \frac{1}{2})$ \\
    $\Pi_{0_{\bm 2}} $ & $ (\ell +\frac{1}{2}) \oplus (\ell - \frac{1}{2})$ \\
    $\Pi_+ $ & $(\ell +1) \oplus \ell $ \\
    $\Pi_- $ & $(\ell - 1) \oplus \ell$ 
\end{tabular}
\caption{Projections $\Pi_k$ and the representations occurring in \eqref{gp} to which they project.}
\vspace{-0.4cm}
\label{table:t2}
\end{center}
\end{table}

A suitable set for the $14$ rotational invariants is provided by the following set of anti-Hermitian matrices
\beqa
&& Q_{0_{\bm 0}}= \Pi_{0_{\bm 0}}Q_B \,, \quad  Q_{0_{\bm 2}}= \Pi_{0_{\bm 2}}Q_B \,, \quad Q_{S1} \,, \quad  Q_{S2} \,, \quad \Pi_{0_{\bm 0}}\,, \quad \Pi_{0_{\bm 2}}\,, \quad \Pi_+ \,, \quad \Pi_- \,,\,  i S_1 \,, \, i S_2 \nn \\
&&Q_- = \frac{1}{4 \ell (\ell +1)}  \Pi_- \left ( (2 \ell+1)^2 Q_B  - i \right ) \Pi_-  \,, \quad Q_+ = \frac{1}{4 \ell (\ell +1)}  \Pi_+ \left ( (2 \ell+1)^2 Q_B  - i \right ) \Pi_+ \nn \\
&&Q_F = {\bm 1}_{2 \ell +1} \otimes \Gamma_a^0 \otimes \sigma_a - i \frac{1}{2} \Pi_{\frac{1}{2}} \,, \quad Q_H = -i \varepsilon_{abc}\frac{X_a \otimes \Gamma_b^0 \otimes \sigma_c}{\sqrt{\ell (\ell +1)}} - \frac{1}{2} Q_{BI} + i\frac{1}{2} \Pi_{\frac{1}{2}} \,, 
\label{eq:rotinvs}
\eeqa
where
\be
Q_B = \frac{X_a \otimes {\bm 1}_4 \otimes \sigma_a - \frac{i}{2}{\bm 1}}{\ell + \frac{1}{2}} \,, \quad Q_{S(i)} = \frac{X_a \otimes s_i \otimes \sigma_a - \frac{i}{2} S_i}{\ell + \frac{1}{2}} \,, \quad  Q_{BI} = i \frac{(\ell + \frac{1}{2})^2 \lbrace Q_B \,, Q_I \rbrace + \frac{1}{2} \Pi_{\frac{1}{2}}}{ 2 \ell (\ell +1)} \,,
\ee
and 
\be
S_i = {\bm 1}_{2 \ell+1} \otimes s_i \otimes {\bm 1}_2 \,, \quad 
s_i =
\left ( 
\begin{array}{cc}
\sigma_i & 0_2 \\
0_2 & 0_2
\end{array}
\right ) \,, \quad i = 1 \,, 2 \,.
\ee

All of these invariants\footnote{We can certainly form a rotational invariant of the natural form $\sigma_a (X_a +\Gamma_a) =\sigma_a D_a$. We note however that this is not linearly independent from the given set of rotational invariants in (\ref{eq:rotinvs}).} are in the matrix algebra $\mbox{Mat}((2 \ell +1) \times 4 \times 2)$. It can be verified that they all commute with $\omega_a$ and that they are linearly independent, so they form a basis for the rotational invariants of $\omega_a$. This is in general not an orthogonal basis under the inner product defined by the $ {\cal N}^{-1} \mbox{Tr}$, although some pairs happen to be orthogonal. It is possible to show that, they square as follows:
\be
\begin{aligned}
Q_B^2 = -1  \,, \quad Q_\pm^2 = - \Pi_\pm \,, \quad Q_{00}^2 = - \Pi_{0_{\bm 0}} \,, \quad Q_{02}^2 = - \Pi_{0_{\bm 2}}\,, \quad Q_{S(i)}^2 = - \Pi_0 \,, \\
(iS_i)^2 = - \Pi_0 \,, \quad Q_F^2 = - \Pi_{\frac{1}{2}} \,, \quad Q_I^2 = - \Pi_{\frac{1}{2}}  \,, \quad Q_{BI}^2 = - \Pi_{\frac{1}{2}} \,, \quad Q_H^2 = -\Pi_{\frac{1}{2}} \,,
\end{aligned}
\ee
from which we observe that, all $iQ$ and $S_i$ are idempotents in the subspaces defined by the relevant projections. It is also easy to observe that 
\be
\begin{aligned}
\Pi_{\frac{1}{2}}  Q_F &= Q_F \,, \quad \Pi_{\frac{1}{2}}  Q_I = Q_I \,, \quad \Pi_{\frac{1}{2}}  Q_H = Q_H \,, \quad \Pi_{\frac{1}{2}}  Q_{BI} = Q_{BI}  \,, \\ 
\Pi_{\frac{1}{2}}  Q_\pm &= Q_\pm \,, \quad \Pi_{\frac{1}{2}}  Q_{S(i)} = 0 \,, \quad \Pi_{\frac{1}{2}} Q_{0_{\bm 0}}= 0 \,, \quad \Pi_{\frac{1}{2}} Q_{0_{\bm 2}}=0 \,.
\end{aligned}
\ee

Using the rotational invariants listed in (\ref{eq:rotinvs}), it is possible to give a suitable basis for the objects that transform as vectors under the adjoint action of $\omega_a$. From \eqref{cgse1} we see that there are $30$ of them and the set of basis vectors for these can be picked as follows:
\be
\begin{aligned}
&\lbrack D_a \,, Q_{0_{\bm 0}}\rbrack \,, \quad Q_{0_{\bm 0}}\lbrack D_a \,, Q_{0_{\bm 0}}\rbrack \,, \quad \lbrace D_a \,, Q_{0_{\bm 0}}\rbrace \,, \\[0.4em]
&\lbrack D_a \,, Q_{0_{\bm 2}}\rbrack \,, \quad Q_{0_{\bm 2}}\lbrack D_a \,, Q_{0_{\bm 2}}\rbrack \,, \quad \lbrace D_a \,, Q_{0_{\bm 2}}\rbrace \,, \\[0.4em]
&\lbrack D_a \,, Q_- \rbrack \,, \quad Q_- \lbrack D_a \,, Q_- \rbrack \,, \quad \lbrace D_a \,, Q_- \rbrace \,, \\[0.4em]
&\lbrack D_a \,, Q_+ \rbrack \,, \quad Q_+ \lbrack D_a \,, Q_+ \rbrack \,, \quad \lbrace D_a \,, Q_+\rbrace \,, \\[0.4em]
&\lbrack D_a \,, Q_H \rbrack \,, \quad Q_H \lbrack D_a \,, Q_H \rbrack \,, \quad \lbrace D_a \,, Q_H\rbrace \,, \\[0.4em]
&\lbrack D_a \,, Q_{F} \rbrack \,, \quad Q_{F} \lbrack D_a \,, Q_{F} \rbrack \,, \quad \lbrace D_a \,, Q_{F} \rbrace \,,  \\[0.4em]
&\lbrack D_a \,, Q_{S1} \rbrack \,, \quad Q_{0} \lbrack D_a \,, Q_{S1} \rbrack \,, \quad \lbrace D_a \,, Q_{S1} \rbrace \,,  \\[0.4em]
&\lbrack D_a \,, Q_{S2} \rbrack \,, \quad Q_{0} \lbrack D_a \,, Q_{S2} \rbrack \,, \quad \lbrace D_a \,, Q_{S2} \rbrace \,,  \\[0.4em]
&\Pi_{0_{\bm 0}}\omega_a \,, \quad  \Pi_{02}\omega_a \,,  \quad \Pi_- \omega_a \,,  \quad  \Pi_+ \omega_a \,, \quad S_1 \omega_a \,, \quad S_2 \omega_a \,.
\end{aligned}
\label{eq:eqvvec}
\ee
where $Q_0= Q_{0_{\bm 0}}+Q_{0_{\bm 2}}= \Pi_0 Q_B$.

Equivariant spinors may be constructed from $\beta_\alpha := {\bf 1}_{2 \ell + 1} \otimes b_\alpha \otimes {\bf 1}_2$ and the rotational invariants given in (\ref{eq:rotinvs}). A linearly independent set of $24$ spinors is provided by the list below:
\be
\begin{aligned}
&\Pi_{0_{\bm 0}}\beta_\alpha Q_- \,, \quad Q_{0_{\bm 0}}\beta_\alpha \Pi_- \,, \quad Q_{0_{\bm 0}}\beta_\alpha Q_- \,, \\[0.4em]
&\Pi_{0_{\bm 0}}\beta_\alpha Q_+ \,, \quad Q_{0_{\bm 0}}\beta_\alpha \Pi_+ \,, \quad Q_{0_{\bm 0}}\beta_\alpha Q_+ \,, \\[0.4em]
&\Pi_- \beta_\alpha Q_{0_{\bm 2}}\,, \quad Q_- \beta_\alpha \Pi_{0_{\bm 2}}\,, \quad Q_- \beta_\alpha Q_{0_{\bm 2}}\,, \\[0.4em]
&\Pi_+ \beta_\alpha Q_{0_{\bm 2}}\,, \quad Q_+ \beta_\alpha \Pi_{0_{\bm 2}}\,, \quad Q_+ \beta_\alpha Q_{0_{\bm 2}}\,, \\[0.4em]
&S_1 \beta_\alpha \Pi_+ \,, \quad S_1 \beta_\alpha \Pi_- \,, \quad \Pi_- \beta_\alpha S_2 \,, \quad  \Pi_+ \beta_\alpha S_2 \,, \\[0.4em]
&Q_{S1} \beta_\alpha \Pi_+ \,, \quad Q_{S1} \beta_\alpha \Pi_- \,, \quad \Pi_- \beta_\alpha Q_{S2} \,, \quad  \Pi_+ \beta_\alpha Q_{S2} 
\,, \\[0.4em]
&Q_{S1} \beta_\alpha Q_+ \,, \quad Q_{S1} \beta_\alpha Q_- \,, \quad Q_- \beta_\alpha Q_{S2} \,, \quad  Q_+ \beta_\alpha Q_{S2} 
\,, \\[0.4em]
\end{aligned}
\label{eq:eqvspi}
\ee

Let us also note that, upon using
\be
\Pi_{\frac{1}{2}} \beta_\alpha \Pi_{\frac{1}{2}} = 0 \,, \quad \Pi_{\frac{1}{2}} \beta_\alpha^\dagger \Pi_{\frac{1}{2}} = 0 \,
\ee
and $\Pi_0 \Pi_{\frac{1}{2}} = 0$, it is readily observed that projection to the $\Pi_{\frac{1}{2}}$ sector leaves all the equivariant spinors projected away. This is naturally expected as no spin $\frac{1}{2}$ representation appears in the Clebsch-Gordan expansion \eqref{cgse1} then.

\subsection{Equivariant Fields in the Monopole Sector}

Projection of the equivariant quantities over $S_F^{2 Int}$ to the monopole sector $S_F^{2 \pm}$ introduced in (\ref{eq:Mon1}) is facilitated by the projectors
\be
(1 - \Pi_\mp)(1 - \Pi_{0_{\bm 2}}) = \Pi_{0_{\bm 2}} + \Pi_\pm \,.
\label{eq:promon1}
\ee
After this projection there are $4$ equivariant scalars, $6$ spinors and $8$ vectors which are given by the following subsets of (\ref{eq:rotinvs}),(\ref{eq:eqvspi}),(\ref{eq:eqvvec}), respectively 
\be
Q_{0_{\bm 0}}\,, \quad Q_\pm \,,  \quad \Pi_{0_{\bm 0}}\,, \quad \Pi_\pm \,, 
\ee
\be
\Pi_{0_{\bm 0}}\beta_\alpha Q_\pm \,, \quad Q_{0_{\bm 0}}\beta_\alpha \Pi_\pm \,, \quad Q_{0_{\bm 0}}\beta_\alpha Q_\pm \,, \quad 
\Pi_\pm \beta_\alpha S_2 \,, \quad \Pi_\pm \beta_\alpha Q_{S2} \,, \quad Q_\pm \beta_\alpha Q_{S2}
\ee
\be
\begin{aligned}
&\lbrack D_a \,, Q_{0_{\bm 0}}\rbrack \,, \quad Q_{0_{\bm 0}}\lbrack D_a \,, Q_{0_{\bm 0}}\rbrack \,, \quad \lbrace D_a \,, Q_{0_{\bm 0}}\rbrace \,, \quad \Pi_{0_{\bm 0}}\omega_a \,,  \\[0.4em]
&\lbrack D_a \,, Q_\pm \rbrack \,, \quad Q_\pm \lbrack D_a \,, Q_\pm \rbrack \,, \quad \lbrace D_a \,, Q_\pm \rbrace \,, \quad \Pi_\pm \omega_a \,.
\end{aligned}
\label{eq:eqvvecp}
\ee
Replacing the $(1 - \Pi_{0_{\bm 2}})$ factor in the projection (\ref{eq:promon1}) with $(1 - \Pi_{0_{\bm 0}})$ leads to an equivalent set of equivariant objects as listed above in which $(\Pi_{0_{\bm 0}}\,, Q_{00})$ is replaced with $(\Pi_{0_{\bm 2}}\,, Q_{02})$.

We can parametrize $A_\mu$ as 
\be
A_\mu = \frac{1}{2} a_\mu^1 Q_{0_{\bm 0}}+  \frac{1}{2} a_\mu^2 Q_\pm +  \frac{1}{2} a_\mu^3 \Pi_{0_{\bm 0}}+  \frac{1}{2} a_\mu^4 \Pi_\pm \,,
\label{eq:eqpar1}
\ee
where $a_\mu^i$ $(i = 1 \,, \cdots \,, 4)$ are $4$ Hermitian gauge fields over the manifold ${\cal M}$. This suggests that we can in general expect to get a $U(1)^{\otimes 4}$ gauge theory after tracing over $S_F^{2 \, \pm}$, unless one or more of the gauge fields decouple from the rest the theory, which could in principal happen at least in the large $\ell$ limit.

Parameterization of $A_a$ in this sector may also be given. It reads
\be
\begin{aligned}A_a =  \frac{1}{2} \varphi_1 \lbrack D_a \,, Q_{0_{\bm 0}}\rbrack +  \frac{1}{2} (\varphi_2 - 1) Q_{0_{\bm 0}}\lbrack D_a \,, Q_{0_{\bm 0}}\rbrack +  i \frac{1}{4 \ell} \varphi_3  \lbrace D_a \,, Q_{0_{\bm 0}}\rbrace +  \frac{1}{2 \ell} \varphi_4 \Pi_{0_{\bm 0}}\omega_a \\
+ \frac{1}{2} \chi_1 \lbrack D_a \,, Q_\pm \rbrack + \frac{1}{2} (\chi_2 - 1) Q_\pm \lbrack D_a \,, Q_\pm \rbrack +  i \frac{1}{4 \ell} \chi_3  \lbrace D_a \,, Q_\pm \rbrace +  \frac{1}{2 \ell} \chi_4 \Pi_\pm \omega_a \,.
\end{aligned}
\ee
where $\varphi_i$ and $\chi_i$ $i = 1 \,, \cdots \, 4$ are real scalar fields over ${\cal M}$.

As $(\Pi_{0_{\bm 0}}\,, Q_{00})$ and $(\Pi_{\pm} \,, Q_{\pm})$ form mutually orthogonal sets under the matrix product, we can save a lot of labor by making contact to our earlier work \cite{Seckin-Derek} and immediately infer the low energy effective action that emerges from this parameterization of the fields as two separate $U(1) \times U(1)$ abelian gauge theories decoupled from each other.\footnote{This is, however not so for models that will emerge from the full sector and also from some other sectors discussed in the next subsection. See the brief remark after (\ref{eq:pm12}).} In the first subspace there are $(a_\mu^1 \,, a_\mu^3)$ as the abelian gauge fields, a complex scalar $\varphi = \varphi_1 + i \varphi_2$ charged under $a_\mu^1$ and two real scalars $\varphi_3$ and $\varphi_4$. Scalar fields $\varphi$,$\varphi_3$ and $\varphi_4$ interact with a quartic potential of the form given in \cite{Seckin-Derek} which reads in the $\ell \rightarrow \infty$ limit
\be
V = \frac{1}{2} (|\varphi|^2 +\varphi_3 -1)^2 +\varphi_3 |\varphi|^2 + \frac{1}{2} \varphi_4^2 \,.
\ee
In the second subspace $(a_\mu^2 \,, a_\mu^4)$ are the abelian gauge fields, complex field $\chi = \chi_1 + i \chi_2$ is charged under $a_\mu^2$ and additionally there are two real scalars $\chi_3$ and $\chi_4$. The interaction potential has the same form as the one above with the substitution $\varphi \rightarrow \chi$. Structure of these two mutually independent sectors are essentially identical; it only differs by the level of the fuzzy sphere corresponding to each sector; $\ell$ and $\ell \pm \frac{1}{2}$, respectively. Abelian Higgs type models mentioned above have attractive and repulsive multi-vortex solutions, which are studied in \cite{Seckin-Derek}.

\subsection{Other Sectors}

We can think of projecting to several other sectors of the full theory. Projection of either of the singlets using $(1 - \Pi_{0_{\bm 2}})$ or $(1 - \Pi_{0_{\bm 0}})$, leads to $8$ scalars, $12$ spinors and $18$ vectors. These may be seen as the equivariant fields of the $U(2)$ theory over the fuzzy spheres
\be
S_F^2 (\ell) \oplus  S_F^2 \left ( \ell + \frac{1}{2} \right ) \oplus S_F^2 \left ( \ell - \frac{1}{2} \right ) \,.
\ee
Scalars are $Q_\pm \,, \Pi_{\pm} \,, Q_F \,, Q_H$ and $(\Pi_{0_{\bm 0}}\,, Q_{0_{\bm 0}})$ or $(\Pi_{0_{\bm 2}}\,, Q_{0_{\bm 2}})$, respectively and spinors and vectors are easily identified from the lists given in (\ref{eq:eqvspi}) and (\ref{eq:eqvvec}).

Projecting away both of the singlet sectors using $(1 - \Pi_0) = (1 - \Pi_{0_{\bm 2}})(1 - \Pi_{0_{\bm 0}})$, i.e. projecting on to the $\Pi_{\frac{1}{2}}$ sectors leaves $6$ equivariant scalars and $14$ equivariant vectors, and no spinors as noted earlier. These may be seen as equivariant fields of the $U(2)$ theory over the space
\be
S_F^2 \left ( \ell + \frac{1}{2} \right ) \oplus S_F^2 \left ( \ell - \frac{1}{2} \right ) \,,
\label{eq:pm12}
\ee
which may be interpreted as a fuzzy monopole bundle of winding number $2$.

It may be useful to consider the parameterizations for the fields $A_\mu$ and $A_a$  for these cases, or for that matter for the full set of equivariants. We may expect that the emerging LEAs will in general be more complicated abelian Higgs type models with several abelian gauge fields, some of which may decouple in the large $\ell$ limit, nevertheless we do not expect that they will all separate into a number of abelian Higgs type models with $U(1) \times U(1)$ gauge symmetry, since in these cases not all the equivariants are mutually orthogonal and many more coupling terms could be foreseen to occur after tracing over the fuzzy spheres. 

Projecting away the $\Pi_{\frac{1}{2}}$ sectors leaves $8$ scalars and $16$ vectors and no spinors. These may be seen as equivariant fields of the $U(2)$ theory over the sector
\be
S_F^2  (\ell) \oplus S_F^2  (\ell) \,.
\ee
In this case the $8$ equivariant scalars are $Q_{00}$, $Q_{02}$, $Q_{S1}$, $Q_{S2}$, $\Pi_{00}$, $\Pi_{02}$, $iS_1$ and $iS_2$. We may view these $Q$ as obtained from
\begin{equation}
{\cal Q} = \left(
    \begin{array}{r@{}c|c@{}l}
  &    \begin{smallmatrix}
        Q & Q \\
        Q &  Q \rule[-1ex]{0pt}{2ex}
      \end{smallmatrix} & \mbox{\Large0}       \\\hline
  &    \mbox{\Large 0} &  \mbox{\Large 0}
           \end{array} 
\right) \,, \quad Q = \frac{X_a \otimes \sigma_a - \frac{i}{2}{\bm1}}{\ell +\frac{1}{2}} \,.
\end{equation}
We then have,
\begin{gather}
Q_{0_{\bm 0}}= \Pi_{0_{\bm 0}}{\cal Q} \Pi_{0_{\bm 0}}\,, \quad Q_{0_{\bm 2}}= \Pi_{0_{\bm 2}}{\cal Q} \Pi_{0_{\bm 2}}\,, \nn \\
Q_{S1} =  \Pi_{0_{\bm 0}}{\cal Q} \Pi_{0_{\bm 2}}+  \Pi_{0_{\bm 2}}{\cal Q} \Pi_{0_{\bm 0}}\,, \quad Q_{S2} = - i \Pi_{0_{\bm 0}}{\cal Q} \Pi_{0_{\bm 2}}+ i \Pi_{0_{\bm 2}}{\cal Q} \Pi_{0_{\bm 0}}\,.
\end{gather}
LEA for this model should involve four decoupled $U(1) \times U(1)$ gauge theories of the type mentioned in the previous section as can be readily inferred from the foregoing discussion. 

\section{Models with $k$-component Multiplets}

We now consider replacing the doublet field $\Psi$ in (\ref{eq:isospinor2}) by a $k$-component multiplet $(k \geq 2)$ of the form
\be
\Psi = \left (
\begin{array}{c}
\Psi_1 \\
\Psi_2 \\
\vdots \\
\Psi_k
\end{array}
\right ) \,,
\label{eq:isospinork}
\ee
of the global $SU(2)$ and $\Psi_\alpha \in Mat({\cal N})$ ($\alpha= 1,2, \,, \cdots k$) are ${\rm SU(}{\cal N}{\rm)}$ scalars transforming under its adjoint representation as $\Psi_\alpha \rightarrow U^\dagger \Psi_\alpha U$. We have 
\be
\Gamma_a = -\frac{i}{2} \Psi^\dagger {\tilde \lambda}_a \Psi \,, \quad {\tilde \lambda}_a = \lambda_a \otimes 1_{\cal N} \,,
\ee
with $\lambda_a$ being the spin $\frac{k-1}{2}$ IRR of $SU(2)$ with $\lbrack \lambda_a \,, \lambda_b \rbrack = 2 i \varepsilon_{abc} \lambda_c$. Under ${\rm SU(}{\cal N}{\rm)}$ these $\Gamma_a$ transforms adjointly as 
\be
\Gamma_a \rightarrow U^\dagger \Gamma_a U \,, \quad U \in {\rm SU(}{\cal N}{\rm )} \,,
\ee
by construction.

Following the line of developments of section 2.2., we see that possible vacuum solutions of the model in the form of direct sums of fuzzy spheres are characterized by the structure of solutions for $\Gamma_a$ satisfying the $SU(2)$ commutation relations 
\be
\lbrack \Gamma_a \,, \Gamma_b \rbrack = \varepsilon_{abc} \Gamma_c \,.
\ee 
To construct these solutions, we introduce $k$ sets of fermionic annihilation-creation operators, fulfilling 
\be
\lbrace b_\alpha \,, b_\beta \rbrace = 0 \,, \quad \lbrace b_\alpha^\dagger \,, b_\beta^\dagger \rbrace = 0 \,, \quad \lbrace b_\alpha \,, b_\beta^\dagger  \rbrace = \delta_{\alpha \beta} \,, \quad \alpha \,, \beta : 1 \,, \cdots \,, k \,.
\ee
They span the $2^k$-dimensional Hilbert space with the basis vectors
\be
|n_1 \,, n_2 \,, \cdots n_k \rangle \equiv (b_1^\dagger)^{n_1} (b_2^\dagger)^{n_2} \cdots (b_k^\dagger)^{n_k} | 0 \,, 0 \rangle \,, \quad n_1 \,, n_2\,, \quad \cdots n_k \in {\mathbb Z}_2 \,. 
\label{eq:basis2}
\ee
Number operator $N = b_\alpha^\dagger b_\alpha$ is valued in the range from $0$ to $k$. ${k \choose n} = \frac{k!}{n! (k-n)!}$ is the number of states with the occupation number $n$ and the total number of $2^k$-states is given by $2^k = \sum_{n=0}^k {k \choose n}$.

Taking the $k$-component multiplet
\be
\psi = \left (
\begin{array}{c}
\psi_1 \\
\psi_2 \\
\vdots \\
\psi_k
\end{array}
\right )
:=
\left (
\begin{array}{c}
b_1 \\
b_2 \\
\vdots \\
b_k
\end{array}
\right ) \,,
\ee
it is easily seen that $\Gamma_a$'s fulfilling the $SU(2)$ commutation relations are given by the $2^k \times 2^k$ matrices
\be
\Gamma_a^0 = -\frac{i}{2} \psi^\dagger \lambda_a \psi \,, \quad \lbrack \Gamma_a^0 \,, N \rbrack = 0 \,,
\ee
and $b_\alpha$, $b_\alpha^\dagger$ satisfy the commutation relations
\be
\lbrack b_\alpha \,, \Gamma_a^0 \rbrack = - \frac{i}{2} (\lambda_a)_{\alpha \beta} b_\beta \,, \quad \lbrack b_\alpha^\dagger \,, \Gamma_a^0 \rbrack =  \frac{i}{2} (\lambda_a)_{\beta \alpha} b_\beta^\dagger \,,
\ee
$\Gamma_a^0$'s form a reducible representation of ${\rm SU(2)}$. To give the IRR decomposition of $\Gamma_a^0$'s we note that all $\Gamma_a^0$ commute with $N$. Therefore the states with a fixed eigenvalue of $N$ forms an IRR of $SU(2)$ and the number of states at a fixed eigenvalue of $N$ corresponds to the dimension of this IRR. Hence, IRRs of $SU(2)$ occurring in the decomposition of $\Gamma_a^0$ may be labeled as
\be
\ell_n^k := \frac{{k \choose n}-1}{2} \,,
\ee
with $n$ denoting the eigenvalue of $N$. What remains is to determine the multiplicities of these representations in the decomposition. Since ${k \choose n} = {k \choose k-n}$, we see that for odd $k$ each IRR appears twice, while for even $k$ each IRR occurs twice except the largest representation, which occurs only once. This happens since ${k \choose \frac{k}{2}} = {k \choose k- \frac{k}{2}}$ holds identically for even $k$. Putting these facts together we can write the IRR content of $\Gamma_a^0$ as 
\be
\begin{aligned}
L_{k \, odd} &:= \ell_0^k \oplus \ell_1^k \oplus \cdots \oplus \ell_k^k = {\bm 2} \sum_{n=0}^{\frac{k-1}{2}} \oplus \ell_n^k \,, \quad \quad \quad \quad \quad \quad  \quad \quad \quad \, \, \, k \, \, \mbox{odd}  \,, \\
L_{k \,even} &:= \ell_0^k \oplus \ell_1^k \oplus \cdots \oplus  \ell_{\frac{k}{2}}^k \oplus \cdots \oplus \ell_k^k = \ell_{\frac{k}{2}}^k \oplus {\bm 2} \sum_{n = 0}^{\frac{k}{2} - 1} \oplus \ell_n^k \,, \quad \quad  k \, \, \mbox{even} \,, 
\end{aligned}
\ee
where $\ell_0^k =  \ell_k^k = 0$, i.e. they are the trivial representations.

If we assume that the dimension ${\cal N}$ of the matrices $\Phi_a$ factorizes as $2^k (2 \ell +1) m$, then the vacuum configurations of the $SU({\cal N})$ gauge theory may be given as
\be
\Phi_a = ( X_a^{(2 \ell + 1)} \otimes {\bf 1}_{2^k} \otimes {\bf 1}_m ) + ( {\bf 1}_{2 \ell + 1} \otimes \Gamma_a^0 \otimes {\bf 1}_m )  \,,
\label{eq:minisol2}
\ee
up to gauge transformations.

Configuration in (\ref{eq:minisol2}) spontaneously breaks the ${\rm U}({\cal N})$ down to ${\rm U}(m)$ which is the commutant of $\Phi_a$ in (\ref{eq:minisol2}). 

$SU(2)$ IRR content of this solution follows from the Clebsch-Gordan decompositions
\be
\begin{aligned}
\ell \otimes L_{k \, odd} &= \sum_{n=0}^{\frac{k-1}{2}} {\bm 2} (\ell + \ell_n^k) \oplus \cdots \oplus {\bm 2} |\ell - \ell_n^k| \,, \\
\ell \otimes L_{k \,even} &= (\ell + \ell_\frac{k}{2}^k) \oplus \cdots \oplus |\ell - \ell_\frac{k}{2}^k| + \sum_{n =0}^{\frac{k}{2} - 1} {\bm 2} (\ell + \ell_n^k) \oplus \cdots \oplus {\bm 2} |\ell - \ell_n^k| \,.
\end{aligned}
\ee
Thus, the vacuum solutions are direct sums of concentric fuzzy spheres
\be
\begin{aligned}
S_{F\,, k \, odd}^{2, Int} &:= \sum_{n=0}^{\frac{k-1}{2}} {\bm 2} S_F^2(\ell + \ell_n^k) \oplus \cdots \oplus {\bm 2} S_F^2 (|\ell - \ell_n^k|) \,, \\
S_{F \,, k \, even}^{2, Int} &:= S_F^2(\ell + \ell_\frac{k}{2}^k) \oplus \cdots \oplus S_F^2(|\ell - \ell_\frac{k}{2}^k|) + \sum_{n = 0}^{\frac{k}{2} - 1} {\bm 2} S_F^2(\ell + \ell_n^k) \oplus \cdots \oplus {\bm 2} S_F^2( |\ell - \ell_n^k|) \,.
\end{aligned}
\label{eq:eofs}
\ee

We see that a particular class of winding number $\pm (k-1)$ monopole bundles are obtained by projecting from $S_{F\,,k \, odd}^{2, Int}$ or $S_{F \,,k \, even}^{2, Int}$ to
\be
S_F^{2 \,, \pm (k-1)} : = S_F^2(\ell) \oplus S_F^2(\ell \pm \ell_1^k) \,.
\ee

Let us look at the cases of $k=3$ and $k=4$ in somewhat more detail. For $k=3$, we have $\Gamma_a$'s carrying the representation ${\bm 2} 0 \oplus {\bm 2} 1$, which is $8$-dimensional. We have
\be
S_{F\,,3}^{2, Int} = {\bm 2} S_F^2(\ell + 1) \oplus {\bm 2} S_F^2(\ell) \oplus{\bm 2} S_F^2(\ell - 1)
\label{eq:ofs}
\ee
and it is possible to show that the adjoint action of the symmetry generators $\omega_a = X_a + \Gamma_a^0 - i \frac{\sigma^a}{2}$ decomposes under Clebsch Gordan series to give $80$-equivariant scalars and $200$ vectors. For $k=4$, $\Gamma_a$'s carry the representation ${\bm 2} 0 \oplus {\bm 2} \frac{3}{2} \oplus \frac{5}{2}$, which is $16$-dimensional. We have
\be
S_{F\,,4}^{2, Int} = {\bm 2} S_F^2(\ell) \oplus S_F^2(\ell + \frac{5}{2}) \oplus {\bm 3} S_F^2(\ell + \frac{3}{2}) \oplus {\bm 3} S_F^2(\ell + \frac{1}{2}) \oplus {\bm 3} S_F^2(\ell - \frac{1}{2}) \oplus {\bm 3} S_F^2(\ell - \frac{3}{2}) \oplus S_F^2(\ell - \frac{5}{2}) \,,
\label{eq:efs}
\ee
In this case, a short calculation yields the number of equivariant scalar, spinors and vectors to be $42$, $24$ and $108$, respectively. 

Another important observation is that, equivariant spinor fields emerge only for even $k$. We can immediately make the consistency of this fact with the equivariance conditions (\ref{eq:cons}) manifest for the $k=3$ case. We see that, the $3$-component multiplet is in the vector representation of the global $SU(2)$ and therefore it transforms as a vector:
\be
\lbrack \omega_a \,, \psi_b \rbrack = \frac{i}{2} (\lambda_a)_{bc} \psi_c = \varepsilon_{abc} \psi_c \,,
\ee
since $(\lambda_a)_{bc} = -2 i  \varepsilon_{abc}$ in the adjoint representation of $SU(2)$.

\section{Connection to the $OSP(2,2)$ and $OSP(2,1)$ Fuzzy Superspheres }

Relation of the vacuum configurations $S_{F}^{2 \,, Int}$ and $S_{F}^{2 \,, \pm}$ to the bosonic (even) parts of $OSP(2,2)$ and $OSP(2,1)$ fuzzy supersphere with $N = 2$ and $N =1$ supersymmetry respectively, may quickly be seen to emerge. Here we follow the references \cite{Book, Balachandran:2002jf}, where a comprehensive discussion of these supergroups and construction of fuzzy superspheres may be found and confine the discussion of their representation theory and associated Lie superalgebras to their pertinent parts that we utilize in this section. 

First, we recall from (\ref{eq:1deco1}) that $S_{F}^{2 \,, Int}$ has the $SU(2)$ IRR content
\be
\left (\ell + \frac{1}{2} \right )\oplus \ell \oplus \ell \oplus \left ( \ell - \frac{1}{2} \right ) \,.
\label{eq:decom1}
\ee

From the representation theory of the supergroup $OSP(2,1)$ it is known that its IRRs are labeled by an integer or half-integer ${\cal J}$, which is called the superspin. This superspin ${\cal J}$ representation of $OSP(2,1)$ decomposes under the $SU(2)$ IRRs as
\be
{\cal J}_{OSP(2,1)} \equiv {\cal J}_{SU(2)} \oplus ({\cal J} - \frac{1}{2})_{SU(2)} \,.
\label{eq:decom2}
\ee
IRRs of $OSP(2,2)$ fall in two categories: typical ${\cal J}_{OSP(2,2)}$ and atypical ${\cal J}^{Atypical}_{OSP(2,2)}$. The latter are irreducible with respect to the $OSP(2,1)$ and in fact they coincide with the superspin ${\cal J}$ representation of $OSP(2,1)$\footnote{For this reason generators $\Lambda_{6,7,8}$ can be non-linearly realized in terms of the generators of $OSP(2,1)$ \cite{Book}.}. Typical representations ${\cal J}_{OSP(2,2)}$ are reducible under the $OSP(2,1)$ IRRs as 
\begin{multline}
{\cal J}_{OSP(2,2)} \equiv {\cal J}_{OSP(2,1)} \oplus ({\cal J} - \frac{1}{2})_{OSP(2,1)}  \\
\equiv {\cal J}_{SU(2)} \oplus ({\cal J} - \frac{1}{2})_{SU(2)} \oplus ({\cal J} - \frac{1}{2})_{SU(2)} \oplus ({\cal J} - 1)_{SU(2)} \,, \quad {\cal J}_{OSP(2,2)} \geq 1 \,,
\label{eq:decom3}
\end{multline}
while $({\frac{1}{2}})_{OSP(2,2)}$ decomposes as
\be
({\frac{1}{2}})_{OSP(2,2)} \equiv ({\frac{1}{2}})_{OSP(2,1)} \oplus (0)_{OSP(2,1)} \equiv ({\frac{1}{2}})_{SU(2)} \oplus (0)_{SU(2)} \oplus (0)_{SU(2)} \,. 
\label{eq:decom4}
\ee

Now, comparing the second line of (\ref{eq:decom3}) with (\ref{eq:decom1}) we see that, they match for ${\cal J}_{OSP(2,2)} = \ell + \frac{1}{2}$. Without going into the details of the construction of fuzzy superspheres, we make the observation that this fact has the immediate implication that $S_{F}^{2 \,, Int}$ is the bosonic part of the $OSP(2,2)$ fuzzy supersphere at superspin level ${\cal J}_{OSP(2,2)} = \ell + \frac{1}{2}$. We also clearly see that the monopole bundles
\be
S_F^{2 \, \pm} := S_F^2 (\ell) \oplus S_F^2 \left ( \ell \pm \frac{1}{2} \right ) \,,
\label{eq:Mon2}
\ee
form the even (bosonic) part of the $OSP(2,1)$ fuzzy supersphere at superspin levels ${\cal J}_{OSP(2,1)} = \ell + \frac{1}{2}$ and ${\cal J}_{OSP(2,1)} = \ell$ for the upper sign and lower sign on (\ref{eq:Mon2}), respectively.

Eight generators of the superalgebra $osp(2,2)$ $\Lambda_i :=(\Lambda_a \,, \Lambda_\mu \,, \Lambda_8)$ $(a=1,2,3) \,, (\mu=4,5,6,7)$ fulfill the graded commutation relations
\be
\begin{aligned}
\lbrack \Lambda_a \,, \Lambda_b \rbrack = i \varepsilon_{abc} \Lambda_c \,, \quad  \lbrack \Lambda_a \,, \Lambda_\mu \rbrack = \frac{1}{2} (\Sigma_a)_{\nu \mu} \Lambda_\nu \,, \quad \lbrack \Lambda_a \,, \Lambda_8 \rbrack = 0 \,, \\
\lbrack \Lambda_8 \,, \Lambda_\mu \rbrack = \Xi_{\mu \nu} \Lambda_\nu \,, \quad \lbrace \Lambda_\mu \,, \Lambda_\nu \rbrace = \frac{1}{2} ({\cal C} \Sigma_a)_{\mu \nu} \Lambda_a + \frac{1}{4} (\Xi {\cal C})_{\mu \nu} \Lambda_8 \,,
\end{aligned}
\ee
where 
\be
\Sigma_a = \left (
\begin{array}{cc}
\sigma_a & 0 \\
0& \sigma_a 
\end{array}
\right ) \,, \quad 
{\cal C } = \left (
\begin{array}{cc}
C & 0 \\
0& -C
\end{array}
\right ) \,, \quad 
\Xi =
\left (
\begin{array}{cc}
0 & I_2 \\
I_2 & 0 
\end{array}
\right ) \,, 
\ee
and $C$ is the two-dimensional Levi-Civita symbol. 

Reality condition on this Lie superalgebra is implemented by the graded dagger operation $\ddagger$, which acts on $\Lambda_a$'s as
\be
\Lambda_a^\ddagger = \Lambda_a^\dagger = \Lambda_a \,, \quad  \Lambda_\mu^\ddagger = - {\cal C}_{\mu \nu} \Lambda_\nu \,, \quad \Lambda_8^\ddagger = \Lambda_8^\dagger = \Lambda_8 \,.
\ee

Restriction to the generators $\Lambda_a$, $(a=1\,\cdots,5)$ give the graded commutation relations of the Lie superalgebra $osp(2,1)$.

It turns out that we can give a construction of the generators of $osp(2,2)$ in the representation $(\frac{1}{2})^{Atypical}_{OSP(2,2)} \equiv (\frac{1}{2})_{OSP(2,1)}$. This is the $3$-dimensional fundamental representation of both $osp(2,2)$ and $osp(2,1)$. 

$\Gamma_a$, $b_\alpha \,, b^\dagger_\alpha$, $N$ and ${\bm 1}_4$ form a basis of $4 \times 4$ matrices acting on the
$4$-dimensional module (\ref{eq:basis1}) carrying the direct sum representation $0_{\bm 0} \oplus 0_{\bm 2} \oplus \frac{1}{2}$ of $su(2)$.
Projecting out the first summand in this direct sum by the projector $(1- P_{0_0})$, we can restrict to the $3$-dimensional  submodule in which we can realize $\Lambda_a$'s as follows
\be
\begin{aligned}
\Lambda_a &: = - i (1- P_{0_0}) \Gamma_a^0 = 
\left ( 
\begin{array}{cc}
0 & 0 \\
0 & \frac{1}{2} \sigma_i
\end{array}
\right ) \,, \quad i = 1,3  \,, \quad 
\Lambda_2 : = i (1- P_{0_0}) \Gamma_2^0 = 
\left ( 
\begin{array}{cc}
0 & 0 \\
0 & \frac{1}{2} \sigma_2
\end{array}
\right ) \,, \\
\vspace{0.3em}
\Lambda_4 &: =  - \frac{1}{2} ({\tilde b}_1 + {\tilde b}_2^\dagger) 
= \frac{1}{2} \left (
\begin{array}{rrr}
0 & 0 & -1 \\
-1 &0 &0 \\
0 & 0 & 0
\end{array}
\right )
\,, \quad \Lambda_5 : = \frac{1}{2} ({\tilde b}_1^\dagger - {\tilde b}_2) 
= \frac{1}{2} \left (
\begin{array}{rrr}
0 & 1 & 0 \\
0 &0 & 0 \\
-1 & 0 & 0
\end{array}
\right ) \,,
\label{eq:f1} \\
\Lambda_6 &: =   \frac{1}{2} ({\tilde b}_1 - {\tilde b}_2^\dagger) 
= \frac{1}{2} \left (
\begin{array}{rrr}
0 & 0 & -1 \\
1 &0 &0 \\
0 & 0 & 0
\end{array}
\right )
\,, \quad \Lambda_7 : = \frac{1}{2} ({\tilde b}_1^\dagger + {\tilde b}_2) 
= \frac{1}{2} \left (
\begin{array}{rrr}
0 & 1 & 0 \\
0 &0 & 0 \\
1 & 0 & 0
\end{array}
\right ) \,, \\
\Lambda_8 &: =  (1- P_{0_0}) N
= \left (
\begin{array}{rrr}
2 & 0 & 0 \\
0 & 1 & 0 \\
0 & 0 & 1
\end{array}
\right ) \,.
\end{aligned}
\ee
In (\ref{eq:f1}) we have introduced the notation
\be
{\tilde b}_\alpha := (1- P_{0_0}) b_\alpha (1- P_{0_0}) \,, \quad {\tilde b}_\alpha^\dagger := (1- P_{0_0}) b_\alpha^\dagger (1- P_{0_0}) \,,
\ee
in which restriction to the $3$-dimensional submodule is understood. For consistency, with the graded dagger operation on $\Lambda_\mu$'s introduced above we have that the graded dagger operation on ${\tilde b}_\alpha$ and ${\tilde b}_\alpha^\dagger$ should be defined as 
\be
{\tilde b}_\alpha^\ddagger = {\tilde b}_\alpha^\dagger \,, \quad ({\tilde b}_\alpha^\dagger)^\ddagger = - {\tilde b}_\alpha \,.
\ee
It can be verified by direct calculation that matrices given in (\ref{eq:f1}) satisfy the commutation relations given above and thereby form the fundamental representation of $osp(2,2)$. Restriction of the matrices to $\Lambda_a$, $(a=1\,\cdots,5)$ gives a realization of the fundamental representation of $osp(2,1)$. 

Let us also note that the $4$-dimensional typical representation $({\frac{1}{2}})_{OSP(2,2)}$ given in (\ref{eq:decom4}) differs from $(\frac{1}{2})^{Atypical}_{OSP(2,2)}$ only by an $SU(2)$ singlet. Keeping the left most column and top most row of zeros after projecting with $(1- P_{0_0})$ in all $\Lambda_a$'s simply gives this $4$-dimensional representation of $OSP(2,2)$.

We find the emergence of these supersymmetry algebras from the vacuum structure of our model intriguing, and although in our model vacuum is purely bosonic, we speculate that perhaps a suitable extension of our model could lead to fuzzy superspheres as their vacuum solution. Our initial attempts along this direction has not been successful;  any progress on this issue will be reported elsewhere. 

\section{Stability of the Vacuum Solutions}

In this section we follow the novel developments and reasoning given in \cite{Acharyya:2014nfa} to argue the stability of vacuum solutions, in the form of direct sums of fuzzy spheres given in (\ref{eq:sfint}). For matrix models, such as the one considered in this paper and also for other string theory related matrix models (for instance those discussed in \cite{Berenstein:2002jq,Dasgupta:2002hx, Myers:1999ps}), potentials may be minimized by choosing the matrix fields as the generators of $su(2)$ Lie algebra, which are in irreducible or reducible representations. For the latter case, vacuum configurations may be seen as forming direct sums of fuzzy spheres, in general. The crucial observation of \cite{Acharyya:2014nfa} is that, such direct sums of fuzzy spheres form mixed states, as long as one or several of the fuzzy spheres at a given level appear more than once in the direct sum, while the vacuum solutions formed by a single fuzzy sphere are pure states \footnote{At this point, it is appropriate to note that the aforementioned developments in \cite{Acharyya:2014nfa} are based on the two recent papers \cite{Balachandran:2012pa, Balachandran:2013kia} addressing in much detail the quantum entropy of mixed states and their associated ambiguities.}. It then follows that, since mixed states can not unitarily evolve to pure states, such vacuum configurations are stable. Following the developments in \cite{Acharyya:2014nfa}, the situation in our case may be understood as follows.  

We have that the matrices $\Phi$ spanning the vacuum configurations treated in this paper are in the matrix algebra ${\cal A} = Mat({\cal N})$. We can consider state $\omega$ on the algebra ${\cal A}$, which is a linear map from ${\cal A}$ to the complex numbers ${\mathbb C}$. This state satisfies
\be
\omega(\Phi^* \Phi)  \geq 0 \,, \quad \forall \Phi \in {\cal A} \,, \quad \omega({\bm 1}) = 1 \,,
\ee 
In this algebraic formalism, a single fuzzy sphere, say at level $L$, may be described by imposing the condition
\be
\omega(X_a X_a ) = L(L+1)  \omega({\bm 1}) = - L(L+1) \,.
\ee
In order to describe direct sums of fuzzy spheres of the form
\be
S_F^{2 \, Int} := S_F^2 (\ell) \oplus S_F^2 (\ell) \oplus  S_F^2 \left ( \ell + \frac{1}{2} \right ) \oplus S_F^2 \left ( \ell - \frac{1}{2} \right ) \,.
\label{eq:sfint3}
\ee
we use the projectors $\Pi_{0_{\bm 0}}$, $\Pi_{0_{\bm 2}}$, $\Pi_+$ and $\Pi_-$, which are of rank $(2 \ell+1)n$, $(2 \ell+1)n$, $(2 \ell +2) n$ and $(2\ell)n$, respectively. Now, we can consider the state $\omega_\alpha$ defined as
\be
\omega_\alpha(\Pi_\alpha {\cal D}_a {\cal D}_a \Pi_\alpha) = - L_\alpha (L_\alpha +1) \,, \quad (\mbox{no sum over $\alpha$})\,,
\label{eq:states}
\ee
where the subscript $\alpha$ take on the values ${0_{\bm 0}}$, ${0_{\bm 2}}$, $+$ and $-$, correspondingly $L_\alpha$ take on the values $\ell$, $\ell$, $\ell+\frac{1}{2}$, $\ell - \frac{1}{2}$, respectively. We recall that the notation ${\cal D}_a$ was introduced earlier in (\ref{eq:calDa}).

The condition introduced by equation (\ref{eq:states}) constrains and splits the matrix algebra ${\cal A}$ into a direct sum of matrix algebras 
\be
{\cal A}_\Pi := Mat((2 \ell+1)n) \oplus Mat((2 \ell+1)n) \oplus Mat((2 \ell+2)n) \oplus Mat((2 \ell)n) \,.
\ee
This corresponds to the decomposition of ${\cal A}$ into the fuzzy spheres in (\ref{eq:sfint3}) where each summand in the latter is tensored with ${\bm 1}_n$.  

Projections corresponding to distinct IRRs are unique up to unitary transformations, while projections corresponding to repeated IRRs are not so. To make this point more concrete, we can first express the projectors $\Pi_\alpha$ in the form\footnote{We note that in the succeding expressions, we write $\alpha$ and $L$ of $L_\alpha$ separately for notational clarity. Thus, we for instance have $|L_\alpha \,, L_3 \rangle = : | L \,, L_{3} \,; \alpha \rangle$.}
\be
\Pi_\alpha = \sum_{L_{3} = -L}^{L} | L \,, L_{3} \,; \alpha \rangle \langle  L\,, L_{3} \,; \alpha | \,, \quad \Pi_\alpha \in {\cal A}_\Pi
\label{eq:propi}
\ee
If we perform a unitary transformation 
\be
| L\,, L_{3} \,; \alpha \rangle = \sum_{\beta} u_{\alpha \beta} |L\,, L_{3} \,; \beta \rangle \,,
\label{eq:trans1}
\ee
where $u \in U(2) \otimes U(1) \otimes U(1)$, then the projectors $\Pi_\alpha$ transform under this unitary transformation as $\Pi_\alpha \rightarrow U^\dagger \Pi_\alpha U$ and take the form 
\be
\Pi_\alpha(u) = \sum_{L_{3} = -L}^{L} \sum_{\beta \,, \gamma} u^\dagger_{\gamma \alpha} u_{\alpha \beta} | L\,, L_{3} \,; \beta \rangle 
\langle  L\,,L_{3} \,; \gamma | \,.
\ee
$\Pi_\alpha(u)$ are still projectors as 
\be
\Pi_\alpha^2 (u) = \Pi_\alpha(u) \,, \quad \Pi_\alpha^\dagger (u) = \Pi_\alpha(u) \,,
\label{eq:prov}
\ee
are easily verified.
 
We note that $u_{\alpha \beta} = \delta_{\alpha \beta}$ for $\alpha \,, \beta = + \,, -$ and therefore $\Pi_\pm(u) = \Pi_\pm$, while the  representations with spin $\ell$ get mixed by the $U(2)$ part of the transformations, i.e. $\Pi_{0_{\bm 0}}(u) \neq \Pi_{0_{\bm 0}}$ and  
$\Pi_{0_{\bm 2}}(u) \neq\Pi_{0_{\bm 2}}$. We see that, although all $\Pi_\alpha$ belongs to ${\cal A}_\Pi$, {\it not} all of the transformed projectors $\Pi_\alpha(u)$ are elements of the algebra of observables ${\cal A}_\Pi$. 

Following \cite{Acharyya:2014nfa}, we can consider the expectation value of an element ${\cal O}$ of ${\cal A}_\Pi$ in the state $\omega$:
\be
\omega({\cal O}) = \sum_\alpha \lambda_\alpha \omega_\alpha({\cal O}) \,, 
\label{eq:expec}
\ee
where $\lambda_\alpha$ is a probability vector ($0 \leq \lambda_\alpha \leq 1$ \,, $\sum_\alpha \lambda_\alpha =1$) and
\be
\omega_\alpha({\cal O}) = \frac{1}{2 L_\alpha +1} \sum_{L_3} \sum_{L_3^\prime} \langle L \,, L_3 \,; \alpha | {\cal O} | L \,, L_3^\prime \,; \alpha \rangle \,.
\label{eq:omegaform}
\ee
It can be checked that, this form of $\omega_\alpha$ is consistent with the condition given in (\ref{eq:states}).

Under the unitary transformation defined by (\ref{eq:trans1}), state $\omega({\cal O})$ remains invariant and therefore we have $U(2) \otimes U(1) \otimes U(1)$ symmetry. It then follows that under the transformation (\ref{eq:trans1})
\be
\lambda_\beta (u) =  \sum_\alpha \lambda_\alpha u^\dagger_{\beta \alpha} u_{\alpha \beta} = \sum_\alpha \lambda_\alpha |u_{\alpha \beta}|^2 \,, \quad \mbox{no sum over $\beta$ in the r.h.s.}
\label{eq:lamuni}
\ee
In accordance with our remarks after (\ref{eq:prov}), under this unitary evolution $\lambda_\pm(u) = \lambda_\pm$, while $\lambda_\alpha(u) \neq \lambda_\alpha$ for $\alpha \neq \pm$ in general.

Passing to the density matrix language, we may express the pure states by the density matrix
\be
\rho_\alpha = |\psi_\alpha \rangle \langle \psi_\alpha| = \sum_{L_{3} \,, L^\prime_3} C^*_{L^\prime_3} C_{L_3} | L \,, L_{3} \,; \alpha \rangle \langle  L\,, L^\prime_{3} \,; \alpha | \,,
\label{eq:density1}
\ee
where
\be
|\psi_\alpha \rangle = \sum_{L_{3}} C_{L_3} | L \,, L_{3} \,; \alpha \rangle \,, \quad \sum_{L_{3}} |C_{L_3}|^2 =1 \,, \quad 0 \leq |C^*_{L^\prime_3} C_{L_3}| \leq 1 \,.
\ee
In view of (\ref{eq:expec}) we also introduce the density matrix $\rho$ as 
\be
\rho = \sum_\alpha \lambda_\alpha(u) \rho_\alpha \,, \quad 0 < \lambda_\alpha < 1 \,,  \quad \sum_\alpha \lambda_\alpha = 1 \,. 
\ee
Expectation values of ${\cal O}$ in the states $\omega_\alpha$ and $\omega$ may now be expressed as 
\be
\omega_\alpha({\cal O}) = Tr (\rho_\alpha {\cal O}) \,, \quad \omega({\cal O}) = Tr (\rho {\cal O}) \,.
\label{eq:traceform}
\ee  
Consistency of $\omega_\alpha$ given in equation (\ref{eq:traceform}) with the equations (\ref{eq:states}) and (\ref{eq:omegaform}) may be easily checked after noting that $\rho_\alpha \Pi_\alpha = \rho_\alpha$.

We observe that the decomposition of $\rho$ into $\rho_\alpha$ as given above is not unique, due to the $U(2) \otimes U(1) \otimes U(1)$ symmetry transforming the $\lambda_\alpha$'s as given in (\ref{eq:lamuni}), therefore $\rho$ is describing a mixed state. This fact may also be seen from
\be
Tr (\rho^2) = \sum_\alpha |\lambda_\alpha(u)|^2  < 1 \,.
\ee
Consequently, the $S_F^{2 Int}$ configuration in equation (\ref{eq:sfint3}) is characterized by the density matrix $\rho$, which is mixed. We conclude, therefore that $S_F^{2 Int}$ is a mixed state. Since a mixed state can not evolve into a pure state under unitary time evolution, decay of $S_F^{2 Int}$ into a single fuzzy sphere $S_F^2$, a pure state, is not possible, hence the $S_F^{2 Int}$ vacuum is stable.

It is possible to say a few words on the von Neumann entropy of $S_F^{2 Int}$. This is given as 
\beqa
S(\rho) &=& - Tr (\rho \log \rho) \nn \\ 
&=& - \sum_\alpha \lambda_\alpha(u) \log \lambda_\alpha(u) + \sum_\alpha \lambda_\alpha(u) S(\rho_\alpha) \,, 
\label{eq:vonNeumann2} \\
&=& - \sum_\alpha \lambda_\alpha(u) \log \lambda_\alpha(u) \nn
\eeqa
where the second line follows from the entropy theorem \cite{Bertlmann} and the third line follows from the fact that $\rho_\alpha$ are pure states and therefore $S(\rho_\alpha)=0$. The transformation in (\ref{eq:lamuni}) is Markovian and since $\sum_\alpha |u_{\alpha \beta}|^2 = \sum_\beta |u_{\alpha \beta}|^2 =1$, it is doubly stochastic. Therefore, the Markov process is irreversible and will increase the entropy of $S_F^{2 Int}$. $S(\rho)$ has the maximal value $S^{max}(\rho) = 2 \log 2$ for $\lambda_\alpha = \frac{1}{4} \,, \forall \alpha$. However, we note that $S^{max}(\rho)$ can only be reached if and only if the system starts with $\lambda_\pm = \frac{1}{4}$, since $\lambda_\pm(u) = \lambda_\pm$. Otherwise, $S(\rho)$ is quenched; it still increases but its maximal value, which is less than $2 \log 2$ is determined by the initial values of $\lambda_\pm$. 

Finally, a similar line of reasoning may be given to show that the vacuum solutions $S_{F\,,k \, odd}^{2, Int}$ and $S_{F\,,k \, even}^{2, Int}$ in (\ref{eq:eofs}) obtained for $k$-component multiplet models are all stable too, as they contain several identical copies of the same $SU(2)$ IRR and therefore they form mixed states. In particular, it is readily observed that the unitary symmetry leading to mixed states for $S_{F\,,k \, 3}^{2, Int}$ in (\ref{eq:ofs}) is $U(2)^{\otimes 3}$, while it is $U(3)^{\otimes 4} \otimes U(2) \otimes U(1) \otimes U(1)$ for $S_{F\,,k \, 4}^{2, Int}$ in (\ref{eq:efs}).

\section{Conclusions and Outlook}

In this work, we have considered an $SU(\cal {N})$ Yang-Mills theory coupled to distinct set of scalar fields which are both in the adjoint representation of $SU({\cal N})$, but forming respectively a doublet and a triplet under the global $SU(2)$ symmetry. We have found that the model spontaneously develops fuzzy extra dimensions, which is given by the direct sum $S_F^{2 \, Int} := S_F^2 (\ell) \oplus S_F^2 (\ell) \oplus  S_F^2 \left ( \ell + \frac{1}{2} \right ) \oplus S_F^2 \left ( \ell - \frac{1}{2} \right )$. We have first examined the fluctuations about the vacuum configuration $S_F^{2 \, Int}$ and reached the result that the spontaneously broken model has the structure a $U(n)$ gauge theory over ${\cal M} \times S_F^{2 \, Int}$. In order to support these results, we have presented complete parameterizations of $SU(2)$-equivariant, scalar, spinor and vector fields characterizing the effective low energy behaviour of the $U(2)$ model on ${\cal M} \times S_F^{2 \, Int}$.  An important outcome of this analysis has been the appearance of equivariant spinor fields, which can be ascribed to  admitting $SU(2)$-doublets (although, implicitly in the form of bilinears) in our model. We have also seen that winding number $\pm 1$ monopole bundles $S_F^{2 \, \pm}$ are naturally contained in $S_F^{2 \, Int}$ and they can be accessed  after certain projections, which we have provided. $SU(2)$-equivariant fields of the $U(2)$ theory over ${\cal M} \times S_F^{2 \, \pm}$ and the low energy features of the latter are also discussed. Introducing $k$-component multiplet of the global $SU(2)$ symmetry into our model we have found new fuzzy extra dimensions that are again given in terms of direct sums of fuzzy spheres, and which also contain a particular class of winding number $\pm (k-1) \in {\mathbb Z}$ monopole bundles $S_F^{2 \,, \pm (k-1)}$. We have also seen that the $SU(2)$-equivariant spinor fields only appear for even $k$ multiplets. Another surprising feature that we have encountered is that $S_F^{2 \, Int}$ identifies with the bosonic part of the $N=2$ fuzzy supersphere with $OSP(2,2)$ supersymmetry. In addition, we were able to construct the generators of the $osp(2,2)$ Lie superalgebra in its 3-dimensional atypical and the 
4-dimensional typical irreducible representation by utilizing the matrix content of the vacuum solution $S_F^{2 \, Int}$.
Finally, we have argued that our vacuum solutions are stable since they form mixed states with non-zero von Neumann entropy.

In a forthcoming publication \cite{Kurkcuoglu-Unal}, we apply our present ideas to $SU({\cal N})$ gauge theories obtained from a massive deformation of the $N=4$ super Yang Mills theory discussed in \cite{Steinacker3}. In addition to scalar fields transforming under the representation $(1,0) \oplus (0,1)$ of the global $SU(2) \times SU(2)$ symmetry, in the same vein to the developments in this paper, we also admit scalar fields in this model transforming under $(\frac{1}{2},0) \oplus (0, \frac{1}{2})$ of the global symmetry, which enter into the action only through their bilinears carrying the $(1,0) \oplus (0,1)$ representation. This model spontaneously develops fuzzy extra dimensions, which may be written as direct sums of the products $S_F^2 \times S_F^2$. In \cite{Kurkcuoglu-Unal} these and related matters will be addressed thoroughly.

\vskip 2em

{\bf \large Acknowledgements}

\vskip 1em

I would like to thank to A.P. Balachandran for many interesting discussions and collaboration at the initial stages of this work. I thank G. \"{U}nal and A. Behtash for several useful discussions and in particular to G. \"{U}nal for help in completing the list of spinor in (\ref{eq:eqvspi}). I also would like to express my thanks to S. Vaidya for an illuminating discussion on the mixed state structure of direct sums of fuzzy spheres. This work is supported by TUBiTAK under project No. 110T738 and TUBA-GEBIP program of The Turkish Academy of Sciences. 

\vskip 1em

\appendices

\section{Identities and Formula related to $\Gamma_a^0$}

Some helpful relations and identities may be listed as follows:
\begin{gather}
P_{\frac{1}{2}} N = N P_{\frac{1}{2}} = P_{\frac{1}{2}} \,, \quad P_{\frac{1}{2}} \Gamma_a^0 = \Gamma_a^0 P_{\frac{1}{2}}  = \Gamma_a^0 \,, \nn \\[0.5em] 
\label{eq:identities2}
(1- P_{0_{\bm 2}}) \Gamma_a^0 = \Gamma_a^0 \,, \quad (1- P_{0_{\bm 2}}) P_{\frac{1}{2}} = P_{\frac{1}{2}} \,,  \quad (1- P_{0_{\bm 2}}) N = P_{\frac{1}{2}} \,, \\[0.5em]
N \Gamma_a^0 = \Gamma_a^0 N = \Gamma_a^0 \,, \quad N^2 = 2 N -  P_{\frac{1}{2}}  \,. \nn  
\end{gather}

Another suitable realization of $\Gamma_a^0$ can be given by introducing the $4 \times 4$ $\gamma$-matrices with the Euclidean signature
\be
\lbrace \gamma_i \,, \gamma_j \rbrace = 2 \delta_{i j} \,.
\ee
Taking
\be
\begin{aligned}
&b_1 = \frac{1}{2} (\gamma_1 + i \gamma_2) \,, \quad b_1^\dagger = \frac{1}{2} (\gamma_1 - i \gamma_2) \,,  \\
&b_2 = \frac{1}{2} (\gamma_3 + i \gamma_4) \,, \quad b_2^\dagger = \frac{1}{2} (\gamma_3 - i \gamma_4) \,,
\end{aligned}
\ee
we can write
\beqa
\Gamma_1^0 &=& - \frac{1}{4} (\gamma_2 \gamma_3 - \gamma_1 \gamma_4) \nn \\
\Gamma_2^0 &=& - \frac{1}{4} (\gamma_1 \gamma_3 + \gamma_2 \gamma_4) \\
\Gamma_3^0 &=& \frac{1}{4} (\gamma_1 \gamma_2 - \gamma_3 \gamma_4) \nn
\eeqa

The associated chirality operator $\gamma_5 = i \gamma_1 \gamma_2 \gamma_3 \gamma_4$ can be expressed in the oscillator realisation as
\be
\gamma_5 = 2 N - 4 N_1 N_2 -1 \,,
\ee
and has the eigenvalue $-1$ on the singlets and $1$ on the doublet. Accordingly the chiral projections are nothing but $P_0$ and $P_{\frac{1}{2}}$ as expected:
\be
P_0 = \frac{(1 - \gamma_5)}{2}  \,, \quad P_{\frac{1}{2}} = \frac{(1 + \gamma_5)}{2}  \,.
\ee

For additional clarity it is useful to have the matrix form of all of these operators in the basis where the rows and columns are given in the order $|0\,,0 \rangle$,$|1 \,,1 \rangle$, $|0\,,1 \rangle$,$|1 \,,0 \rangle$, 
\be
\begin{aligned}
b_1 &\text{:=}\left(
\begin{array}{cccc}
 0 & 0 & 0 & 1 \\
 0 & 0 & 0 & 0 \\
 0 & 1 & 0 & 0 \\
 0 & 0 & 0 & 0 \\
\end{array}
\right)
\,, \quad 
b_2 \text{:=}\left(
\begin{array}{cccc}
 0 & 0 & -1 & 0 \\
 0 & 0 & 0 & 0 \\
 0 & 0 & 0 & 0 \\
 0 & 1 & 0 & 0 \\
\end{array}
\right) \\
\vspace{0.4em}
b_1^{\dagger } & \text{:=}\left(
\begin{array}{cccc}
 0 & 0 & 0 & 0 \\
 0 & 0 & 1 & 0 \\
 0 & 0 & 0 & 0 \\
 1 & 0 & 0 & 0 \\
\end{array}
\right)
\,, \quad 
b_2^{\dagger }\text{:=}\left(
\begin{array}{cccc}
 0 & 0 & 0 & 0 \\
 0 & 0 & 0 & 1 \\
 -1 & 0 & 0 & 0 \\
 0 & 0 & 0 & 0 \\
\end{array}
\right)
\end{aligned}
\ee

\be
N \text{:=}\left(
\begin{array}{cccc}
 0 & 0 & 0 & 0 \\
 0 & 2 & 0 & 0 \\
 0 & 0 & 1 & 0 \\
 0 & 0 & 0 & 1 \\
\end{array}
\right) \,, \quad 
N_1 \text{:=}\left(
\begin{array}{cccc}
 0 & 0 & 0 & 0\\
 0 & 1 & 0 & 0 \\
 0 & 0 & 0 & 0 \\
 0 & 0 & 0 & 1 \\
\end{array} \right) 
\,, \quad
N_2 \text{:=}\left(
\begin{array}{cccc}
 0 & 0 & 0 & 0 \\
 0 & 1 & 0 & 0 \\
 0 & 0 & 1 & 0 \\
 0 & 0 & 0 & 0 \\
\end{array}
\right) \,,
\ee

\be
P_{0_{\bm 0}} \text{:=}\left(
\begin{array}{cccc}
 1 & 0 & 0 & 0 \\
 0 & 0 & 0 & 0 \\
 0 & 0 & 0 & 0 \\
 0 & 0 & 0 & 0 \\
\end{array}
\right) \,, \quad 
P_{0_{\bm 2}} \text{:=}\left(
\begin{array}{cccc}
 0 & 0 & 0 & 0 \\
 0 & 1 & 0 & 0 \\
 0 & 0 & 0 & 0 \\
 0 & 0 & 0 & 0 \\
\end{array}
\right) \,, \quad 
P_{\frac{1}{2}} \text{:=}\left(
\begin{array}{cccc}
 0 & 0 & 0 & 0 \\
 0 & 0 & 0 & 0 \\
 0 & 0 & 1 & 0 \\
 0 & 0 & 0 & 1 \\
\end{array}
\right) \,, \quad 
\ee

\be
\gamma_1 \text{=}\left(
\begin{array}{cc}
 0 & \sigma_1 \\
 \sigma_1 & 0 \\
\end{array}
\right) \,, \quad 
\gamma_2 \text{=}\left(
\begin{array}{cc}
 0 & \sigma_2 \\
 \sigma_2 & 0 \\
\end{array}
\right) \,, \quad 
\gamma_3 \text{=} - \left(
\begin{array}{cc}
 0 & \sigma_3 \\
 \sigma_3 & 0 \\
\end{array}
\right) \,, \quad 
\gamma_4 \text{=} i \left(
\begin{array}{cc}
 0 & {\bm 1}_2 \\
 -{\bm 1}_2 & 0 \\
\end{array}
\right) \,, \quad 
\ee
and 
\be
\gamma_5 \text{=}  \left(
\begin{array}{cc}
- {\bm 1}_2& 0 \\
 0 & {\bm 1}_2 \\
\end{array}
\right) \,,
\ee

\end{document}